\documentstyle[seceq,epsf]{ptptex}

\newcommand{\beq}{\begin{eqnarray}}
\newcommand{\eeq}{\end{eqnarray}}
\newcommand{\btem}{\bibitem}

\newcommand{\bep}{\bar{\varepsilon}}
\newcommand{\vep}{\varepsilon}
\newcommand{\vphi}{\varphi}

\renewcommand{\d}{\partial}
\preprintnumber[5cm]{UTHEP-420, Mar. 2000\\ KUNS-1649}

\title{
Optimized Perturbation Theory at Finite Temperature
}
\subtitle{Two-Loop Analysis}
\author{
Suenori {\sc CHIKU}
}

\inst{
Institute of Physics, University of Tsukuba,
 Tsukuba, Ibaraki 305-8571, Japan\\
 and\\
 Physics Department, Kyoto University, Kyoto 606-8502, Japan
}

\recdate{
February 25, 2000
}
\abst{

 We study the optimized perturbation theory (OPT) at finite 
 temperature, which is a self-consistent resummation method.
 Firstly, we generalize the idea of the OPT to optimize the coupling constant 
 in $\lambda \phi^4$ theory, and give a proof of the renormalizability of 
 this generalized OPT.
 Secondly, the principle of minimal sensitivity and 
 the criterion of the fastest apparent convergence,
 which are conditions to determine the optimal parameter values,
 are examined in $\lambda \phi^4$ theory. Both conditions exhibit
 a second-order transition at finite temperature with critical 
 exponent $\beta = 0.5$ in the two-loop approximation.
}
\begin{document}
\maketitle
\section{Introduction}

 Experiments on ultra-relativistic heavy ion collisions at RHIC 
 (the Relativistic Heavy Ion Collider) and LHC (the Large Hadron Collider)
 are being planned for the purpose of producing hot matter 
 similar to that realized in the early  universe. \cite{QM98}
 In such a high-temperature ($T$) phase, quarks and gluons are deconfined
 and chiral symmetry is expected to be restored. The dynamics of such 
 a phase transition from hadronic matter to a quark-gluon plasma
 is described by quantum chromodynamics (QCD) at finite $T$.

 At present, there are two major calculational methods in QCD at finite $T$.
 One is the numerical simulations in lattice QCD. \cite{LAT98}
 This method has revealed that the restoration of chiral symmetry takes place 
 at $T_c \sim 150$ MeV. For two flavors massless, the phase transition 
 has been shown to be of second order. Analysis of the case with 
 two light flavors and one medium-heavy quark, which corresponds 
 to the real world, is also under study \cite{ukawa}. 
 One of the disadvantages of lattice QCD is the difficulty of treating
 real-time excitations at finite $T$ in a straightforward way 
 on the Euclidean lattice. \cite{nakahara}

 Another method of treating QCD for $T \neq 0$ is based on the hard 
 thermal loops resummation scheme (HTLRS). \cite{BP}
 It is generally known that naive perturbation 
 theory breaks down at high temperature, \cite{WEIN74,KL76}
 even if the coupling constant is small. This is the reason that we must 
 resum higher-order terms at finite $T$.  HTLRS is a method of resuming
 the higher-order terms at high $T$ up to $O(gT)$ 
 ($g$ is the QCD coupling constant) when the external momenta 
 are $q \sim O(gT)$, a situation referred to as `soft'.
 With this method, physical quantities such as the soft gluon damping 
 rate and the dilepton production rate have been calculated 
 in the quark-gluon plasma (QGP) phase. \cite{CHTL}
 Since it is an effective resummation method which works only 
 at high $T$, it cannot be applied to systems at low $T$.
 Also, in theories with spontaneous symmetry breaking (SSB),
 the loop expansion is more relevant than weak-coupling expansions, 
 such as HTLRS. 

 To treat non-perturbative QCD effects at low energies, related to
 spontaneous symmetry breaking, effective models, such as 
 the linear $\sigma$ model, have been developed. In such approaches, however,
 the serious problem of tachyonic poles appear \cite{WEIN74}, 
 when one studies symmetry restoration at finite $T$.
 This problem shows up even below $T_c$, and it causes the breakdown of 
 the thermal perturbation theory. Therefore, a resummation method 
 applicable to a wide range of $T$ from low $T$ to high $T$
 is required when effective models are studied.

 To this time, several methods have been proposed for self-consistent
 resummation. \cite{others} In these methods, there appear two major 
 problems: \cite{renorm}
 (i) appearance of $T$-dependent ultraviolet divergences, 
 which makes the renormalizability non-trivial at finite $T$,
 and (ii) the breakdown of the Nambu-Goldstone (NG) theorem at finite $T$ 
 when SSB occurs. 

 In Ref.\cite{CH98}, we have generalized the optimized perturbation 
 theory (OPT) at finite $T$, which is a resummation method for solving 
 the above mentioned problems. 
 A similar idea has been proposed under the names of
 the delta-expansion, the variational perturbation theory, and so on.
 These ideas are applied to field theory at finite $T$ in 
 Ref.\cite{oko} and \cite{BM}.
 In Ref.\cite{oko}, the $O(N)$ $\phi^4$ theory with 
 the principle of minimal sensitivity (PMS) (see \S
 \ref{sec:OPT}) is studied. Since the formulation in that work 
 is not based on the loop expansion, its authors could not 
 prove renormalizability. 
 Also, the NG-theorem is not satisfied in a straightforward way.
 In Ref.\cite{BM}, $\lambda \phi^4$ theory with
 the fastest apparent convergence (FAC) condition (see \S \ref{sec:OPT}) 
 at high $T$ is studied. There, the renormalization up to 
 the two-loop order is carried out in the symmetric phase. 
 However, a rigorous definition of the method at higher orders and a proof of 
 its renormalizability are not given. 

 Our method has the following advantages over the previous methods.
\begin{enumerate}
	\item Renormalization can be carried out automatically
	at each order of the OPT.
	\item The Nambu-Goldstone theorem is satisfied in any given 
	order of the OPT.
\end{enumerate}

 In this paper, we further generalize the OPT studied in Ref.\cite{CH98} 
 to optimize the coupling constant in the $\lambda \phi^4$ theory.
 Also, we investigate whether the phase transition is correctly
 described by the OPT in $\lambda \phi^4$ theory, which is known to have
 a second order transition. 
 We examine the principle of minimal sensitivity and 
 the criterion of the fastest apparent convergence,
 which are conditions to determine the optimal parameter values,
 in the two-loop approximation.

 The organization of this paper is as follows. 
 In \S \ref{sec:ext}, we generalize the OPT introduced in Ref.\cite{CH98}
 to optimize the coupling constant in $\lambda \phi^4$ theory. 
 The proof of renormalizability in this generalized OPT is also given. 
 In \S \ref{sec:phase}, we investigate the phase transition in 
 $\lambda \phi^4$ theory in the OPT.
 The principle of minimal sensitivity (PMS) and the criterion of 
 the fastest apparent convergence (FAC) \cite{PMS} 
 at the two-loop order are examined.
 In \S \ref{sec:summary}, we give summary and concluding remarks.

\section{Formulation of the OPT with optimization of the coupling constant}
\label{sec:ext}

 In $\lambda \phi^4$ theory with spontaneous symmetry breaking (SSB), 
 the naive perturbation theory has the following difficulties.
\begin{itemize}
	\item Higher-order terms give larger contributions than 
		lower order terms (the existence of the hard thermal loop).
	\item Tachyonic poles appear when one studies the restoration 
		of symmetries.
\end{itemize}
 In the following, we introduce the optimized perturbation theory (OPT) 
 with both mass and coupling constant optimization.
 This theory allows us to avoid the above problems.

\subsection{Formulation of the OPT}
\label{sec:OPT}

 We reformulate the OPT to incorporate the optimization of 
 the coupling constant.
 This extension can be performed in the same way in Ref.\cite{CH98}.
 Let us start with the renormalized Lagrangian ${\cal L}$ of
 $\lambda \phi^4$  theory:
\beq
\label{eq:ala1}
	{\cal L}(\phi;\mu^2,\lambda) 
	& = & 	{1 \over 2} [(\d \phi)^2 
	- \mu^2 \phi^2] -{\lambda \over 4!} \phi^4 \nonumber \\
	&& +{1 \over 2} A(\lambda) (\d \phi)^2 
	- {1 \over 2} B(\lambda) \mu^2 \phi^2
	- {1 \over 4! } C(\lambda)  \phi^4 + D(\lambda) \mu^4 
\eeq
 Here we have explicitly written the arguments $\mu^2$ and $\lambda$
 in ${\cal L}$ for later use. 
 Since we adopt the $\overline{MS}$ scheme \cite{msbar} (which is 
 a mass-independent renormalization scheme) with dimensional 
 regularization, $A$, $B$, $C$ and $D$ are functions
 of $\lambda$ only. 
 For simplicity, we omit the dimensional factor $\kappa^{4-n}$ 
 that multiplies $\lambda$. Here, $\kappa$ is the renormalization
 point and $n$ is the number of dimensions.

\vspace*{0.2cm}
\noindent
{\bf Step 1} (Definition of the $\delta$ expansion)

 The loop-wise $\delta$ expansion for the effective action is defined as
\beq
\label{ep:naive-l}
	{\Gamma[\varphi^2]} = \delta \hbar \ln \int[d \phi]
	\exp \left[ {1 \over \delta \hbar } \int_0^{ \hbar /T} d^4 x
	\left[ {\cal L} (\phi + \varphi;\mu^2, \lambda) + J \phi
	\right] \right] ,	
\eeq
 where $J \equiv - \d \Gamma[\varphi] /\d \varphi $ and
 $\int_0^{ \hbar /T} d^4 x \equiv \int_0^{ \hbar /T} d\tau \int d^3 x$. 
 At $T = 0$, it is equivalent to the naive $\hbar$ 
 expansion. \cite{EFF} However, it does not coincide with 
 the $\hbar$ expansion for finite $T$
 because $\hbar$ is also contained in the upper limit of the integral.

 The counterterms are also expanded in powers of $\delta$. Because 
 the renormalization is performed at $T = 0$, the counterterms are
 the same as those in the naive loop expansion. 

 Note that $\delta$ is introduced only to define the order of 
 the perturbation theory. At the end, we set $\delta = 1$.

\vspace*{0.2cm}
\noindent
{\bf Step 2} (Splitting the mass and coupling)

 The mass and {\it the coupling constant} are decomposed as
\beq
\label{eq:split}
\begin{array}{rcccl}
	\mu^2 & = & m^2 - (m^2 - \mu^2) & = & m^2 - \chi, \\
	\lambda & = & g - (g - \lambda) & = & g - \eta.
\end{array}
\eeq
 Namely, we add and subtract the mass term $m^2$ and the coupling constant
 $g$, and we define $\chi \equiv m^2 - \mu^2$ and $\eta \equiv g - \lambda$.
 Using the expression in Eq.(\ref{eq:split}) to substitute into 
 Eq.(\ref{eq:ala1}), we can rewrite the Lagrangian as
\beq
\label{eq:rela}
	 {\cal L}(\phi;m^2, \chi,g, \eta) 
	& \equiv & {\cal L}(\phi;m^2 - \chi,g - \eta) \nonumber \\
	& = & 	{1 \over 2} [(\d \phi)^2 - m^2 \phi^2] 
	- {g \over 4!} \phi^4  + {1 \over 2} \chi \phi^2 
	+ {\eta \over 4!} \phi^4 \nonumber \\
	&& +{1 \over 2} A(g - \eta) (\d \phi)^2 
	- {1 \over 2} B(g - \eta) (m^2 - \chi) \phi^2 \nonumber \\
	&& - {1 \over 4! } C(g - \eta)  \phi^4 
	+ D (g - \eta) (m^2 - \chi^2)^2. 
\eeq
 It is important that the identities in Eq.(\ref{eq:split}) are used 
 not only in the standard mass and coupling terms but also in
 the counterterms \footnote{This point was first made in Ref.\cite{BM}.} 
 to make the order by order renormalization possible in the OPT.

 To obtain a non-trivial loop expansion, we need to assign $\delta$ as
\beq
\label{eq:oofd}
	m^2=O(1),\ \  \lambda=O(1),\ \ 
	\chi=O(\delta),\ \ \eta=O(\delta).
\eeq
 Thus, the tree-level mass becomes $m^2 + g \vphi^2 /2$ instead of
 $\mu^2 + \lambda \vphi^2 /2$ in the symmetry broken phase. 
 The order of $\delta$ is increased by 
 inserting the new vertex $\chi \phi^2 /2$ or $\eta \phi^4 / 4!$.
 The physical reason behind the relations in Eq.(\ref{eq:oofd}) is 
 the fact that $\chi$ and $\eta$ reflect the effect of interactions.

 Since Eq.(\ref{eq:rela}) is simply a rearrangement of the parameters,
 the effective action should not depend on the arbitrary
 parameters $m^2$ and $g$. However, since we cannot calculate all orders 
 in actual calculations, the physical quantities depend on artificial
 parameters.
 Methods for determination of these parameters are given in the next step.

\vspace*{0.2cm}
\noindent
{\bf Step 3} (Determination of $m^2$ and $g$)

 One can determine the optimal parameter values of $m^2$ and $g$ 
 using methods proposed by Stevenson. \cite{PMS} 
\begin{itemize}
\item[(a)] The principle of minimal sensitivity (PMS): 
\beq
\label{eq:pmsc}
	{\partial {\cal O}_L \over \partial m } = 0, \ \ 
	{\partial {\cal O'}_{L'} \over \partial g } = 0.
\eeq
 ${\cal O}_L$ (${\cal O'}_{L'}$) represents a physical quantity calculated 
 up to $L$-th ($L'$-th) order. Since $m^2$ and $g$ are artificial parameters
 added by hand, the physical quantities should not depend on them.
 \item[(b)] The criterion of the  fastest apparent convergence (FAC): 
\beq
\label{eq:facc}
	{\cal O}_L - {\cal O}_{L-n} = 0, \ \ 
	{\cal O'}_{L'} - {\cal O'}_{L'-n'} = 0,
\eeq
 where $n$ ($n'$) is chosen in the range $1 \le n \le L$ ($1 \le n \le L'$).
 This condition requires that the perturbative corrections in
 ${\cal O'}_L$ (${\cal O'}_{L'}$) should be as small as possible 
 for a suitable value of $m$ ($g$).
\end{itemize}
 These conditions are reduced to self-consistent gap equations.
 Therefore, OPT corresponds to a generalization of the mean field 
 approximation. 

 An extension of the Lagrangian to the $O(N)$ case is straightforward.
 Actually, one can easily see that the Nambu-Goldstone theorem is fulfilled 
 in completely the same way as in Ref.\cite{CH98}.

 In the next section, proof of the  renormalizability in this extended OPT
 is given.
 
\subsection{Proof of the renormalizability in the OPT}
\label{sec:prof-reno}

 Consider a naive $m$-th loop order renormalized correction to
 the effective action, $\Gamma_R^{(m)}$. Then,
 the effective action up to $n$-th order can be written as
\beq
\label{eq:gamma-d}
	\Gamma_R^n = \sum_{m=0}^n \Gamma_R^{(m)}.
\eeq
 Here, each $\Gamma_R^{(m)}$ is written as the sum of $\Gamma^{(m)}$, 
 which does not contain any counterterms, and $C^{(m)}$, which 
 includes more than one counterterms: 
\beq
\label{eq:gamma-s}
	\Gamma_R^{(m)} = \Gamma^{(m)} + C^{(m)}.
\eeq
 Also, $\Gamma_R^{(m)}$ can be expanded in the renormalized mass $\mu^2$ 
 and the renormalized coupling $\lambda$ around $\mu^2 =0$ and $\lambda =0$:
\beq
\label{eq:gamma-e}
	\Gamma_R^{(m)} = \sum_{i=0}^{\infty} \sum_{j=0}^l 
	(\mu^2)^i \lambda^j (\Gamma^{(m)}_{ij} + C^{(m)}_{ij}).
\eeq
 Here
\beq
\label{eq:co-def}
	\Gamma^{(m)}_{ij} & = & \left. {1 \over i! j!}{\d^i \over \d (\mu^2)^i}
	{\d^j \over \d \lambda^j} \Gamma^{(m)} \right|_{\mu^2=0,\lambda=0},\\
	C^{(m)}_{ij} & = & \left. {1 \over i! j!}{\d^i \over \d (\mu^2)^i}
	{\d^j \over \d \lambda^j} C^{(m)} \right|_{\mu^2=0,\lambda=0}.
\eeq
 Since $\Gamma_R^{(m)}$ is finite, $\Gamma^{(m)}_{ij} + C^{(m)}_{ij}$
 must also be finite.

 Next, we consider the OPT. In this method, $\mu^2$ and $\lambda$ are
 rewritten in Eq.(\ref{eq:split}).
 Then, the expressions in Eq.(\ref{eq:split}) are used to substituted 
 into (\ref{eq:gamma-e}), yielding
\beq
\label{eq:gamma-opt}
	\Gamma_R^{(m)} = \sum_{i=0}^{\infty} \sum_{j=0}^l 
	(m^2 - \chi )^i (g- \eta)^j (\Gamma^{(m)}_{ij} + C^{(m)}_{ij}).
\eeq
 Since $m^2$ is $O(1)$, $\lambda$ is $O(1)$, $\chi$ is $O(\delta)$ 
 and $\eta$ is $O(\delta)$,
 the order of $\delta$ increases if terms include
 $\chi$, which is $O(\delta)$ and/or $\eta$, which is $O(\delta)$.
 Thus, by expanding the parameters $(m^2 - \chi )^i$ and $(g- \eta)^j$ 
 in Eq.(\ref{eq:gamma-opt}),terms of higher order in  $\delta$ are generated:
\beq
	\Gamma_R^{(m)} & = & \Gamma_R^{(m)}(\delta^m) 
	+ \Gamma_R^{(m)}(\delta^{(m+1)}) + \cdots \nonumber \\
	& = & \sum_{i=0}^{\infty} \sum_{j=0}^l [ (m^2)^i g^j 
	- \{ i \chi (m^2)^{i-1} g^j + j \eta (m^2)^i g^{j-1}\} + \cdots ]
	(\Gamma^{(m)}_{ij} + C^{(m)}_{ij}), 
\eeq
 where
\beq
	\Gamma_R^{(m)}(\delta^{m+a+b}) \equiv 
	{}_iC_a (-\chi)^a (m^2)^{i-a} {}_jC_b (-\eta)^b g^{j-b}
	(\Gamma^{(m)}_{ij} + C^{(m)}_{ij}).
\eeq
 Thus, up to the $m$-th $\delta$ order correction, 
 $\Gamma_R^{\delta^m}$ reads
\beq
\label{eq:decon}
	\Gamma_R^{\delta^m} = \sum_{s=0}^m \Gamma_R^{(m-s)}(\delta^m).
\eeq
 Since the renormalization is carried out up $m$-th order, each 
 $\Gamma_R^{(l)}$ ($0 \le l \le m$) is finite. 
 Thus, $\Gamma_R^{\delta^m}$ is also finite.

 If a theory does not have a mass term, the renormalization of 
 the OPT is defined as the limit $\mu^2 \rightarrow 0$ of Eq.(\ref{eq:decon}). 
 By using this definition, one can obtain the Debye mass even 
 when the original Lagrangian does not have a mass term.

\section{Phase transition in $\lambda \phi^4$ theory}
\label{sec:phase}

\subsection{Calculation of the effective potential}
\label{sec:caEP}

 In this section, we apply the OPT to $\lambda \phi^4$ theory.
 Firstly, we calculate the effective potential for finite $T$
 up to the 2-loop level using the real time formalism. \cite{real,realeff}
 The coefficients of the counterterms for Eq.(\ref{eq:ala1}) at the 2-loop 
 level \cite{Ramond} are
\beq
\label{eq:cocount}
	A(\lambda) & = & - {\lambda^2 \over (4 \pi)^4}{1 \over 24 \bep}, 
	\nonumber \\
	B(\lambda) & = & {\lambda \over (4 \pi)^2}{1 \over 2 \bep}
	+ {\lambda^2 \over (4 \pi)^4} \left(
	{1 \over 2 \bep^2} - {1 \over 4 \bep} \right), \nonumber \\
	C(\lambda) & = & {\lambda^2 \over (4 \pi)^2}{3 \over 2 \bep}
	+ {\lambda^3 \over (4 \pi)^4} \left(
	{9 \over 4 \bep^2} - {3 \over 2 \bep} \right), \nonumber \\
	D(\lambda) & = & - {1 \over (4 \pi)^2}{1 \over 4 \bep}
	- {\lambda \over (4 \pi)^4}{1 \over 8 \bep^2}, 
\eeq
 where we adopt the $\overline{MS}$ scheme. (The factor 
 $\kappa^{(4-n)}$ multiplying $\lambda$ has been dropped, as above.)
 The effective potential for (\ref{eq:ala1}) is calculated 
 in Ref.\cite{V2T0} at $T = 0$ and in Ref.\cite{realeff}, \cite{PAR} 
 and \cite{Esponosa} for $T \neq 0$. 
 The result is (see Appendix \ref{app:2-loop})
\beq
\label{eq:Vorder}
	V & = & V^0 + V^1 + V^2,\\
\label{eq:Vorder0}
	V^0  & = & {1 \over 2} \mu^2 \vphi^2 +{\lambda \over 4!} \vphi^4,\\
\label{eq:Vorder1}
	V^1  & = & -{1 \over (4 \pi)^2}{M^4 \over 4}
	({3 \over 2}-\ln{M^2 \over \kappa^2}) + \int^{\infty}_0 
	{dk \over (2 \pi)^2} {2 \over \beta}k^2 \ln(1-e^{-\beta E}),\\
\label{eq:Vorder2}
	V^2 & = &  {\lambda \over 2} K_t^2 + {\lambda^2 \vphi^2 \over 4} S_s 
		+ {\lambda^2 \vphi^2 \over 4} C_f,
\eeq
 where $M = \mu^2 + \lambda \vphi^2 / 2$, $\beta = 1 / T$ and 
 $E = \sqrt{k^2 + M^2}$.
 The definitions of $K_t$, $S_s$ and $C_f$ are found 
 in Appendix \ref{app:2-loop} (Eqs.(\ref{eq:defKt}), (\ref{eq:setsun}) 
 and (\ref{eq:defCf}), respectively).

\subsection{Application of the OPT to $\lambda \phi^4$ theory}
\label{sec:appOPT}

 Let us apply the OPT to Eq.(\ref{eq:ala1}). This leads to
\beq
\label{eq:p4OPT}
	{\cal L}(\phi;m^2, \chi,g, \eta) & =  & 
	{1 \over 2} [(\d \phi)^2 - m^2 \phi^2] 
	- {g \over 4!} \phi^4  + {1 \over 2} \chi \phi^2 
	+ {\eta \over 4!} \phi^4 \nonumber \\
	&& +{1 \over 2} A(g - \eta) (\d \phi)^2 
	- {1 \over 2} B(g - \eta) (m^2 - \chi) \phi^2 \nonumber \\
	&& - {1 \over 4! } C(g - \eta)  \phi^4 
	+ D (g - \eta) (m^2 - \chi^2)^2,
\eeq
 where we have used Eq.(\ref{eq:split}). 
 New diagrams for the one-point functions and the vacuum energy diagrams 
 are given in Appendix \ref{app:fullV} in Figs.\ref{fig:atado} 
 \ref{fig:aloop}, respectively.
 
 Up to $O(\delta^2)$, the effective potential Eq.(\ref{eq:Vorder}) becomes
\beq
\label{eq:deffullV}
	V & = & V^{0} + V^{\delta} + V^{\delta^2},\\
\label{eq:fullV0}
	V^{0}  & = & {1 \over 2} m^2 \vphi^2 +{g \over 4!} \vphi^4,\\
\label{eq:fullV1}
	V^{\delta}  & = & -{1 \over 2} \chi \vphi^2 -{\eta \over 4!} \vphi^4
		-{1 \over (4 \pi)^2}{M^4 \over 4}
		({3 \over 2}-\ln{M^2 \over \kappa^2}) \nonumber \\
		&& + \int^{\infty}_0 {dk \over (2 \pi)^2}
		{2 \over \beta}k^2 \ln(1-e^{-\beta E}),\\
\label{eq:fullV2}
	V^{\delta^2} & = & (\chi + {\eta \vphi^2 \over 2} 
		+ {g \over 2} K_t) K_t + {g^2 \vphi^2 \over 4} (S_s + C_f).
\eeq

\begin{figure}[ht]
\centerline{
    \epsfxsize=8.9cm 
    \epsfbox{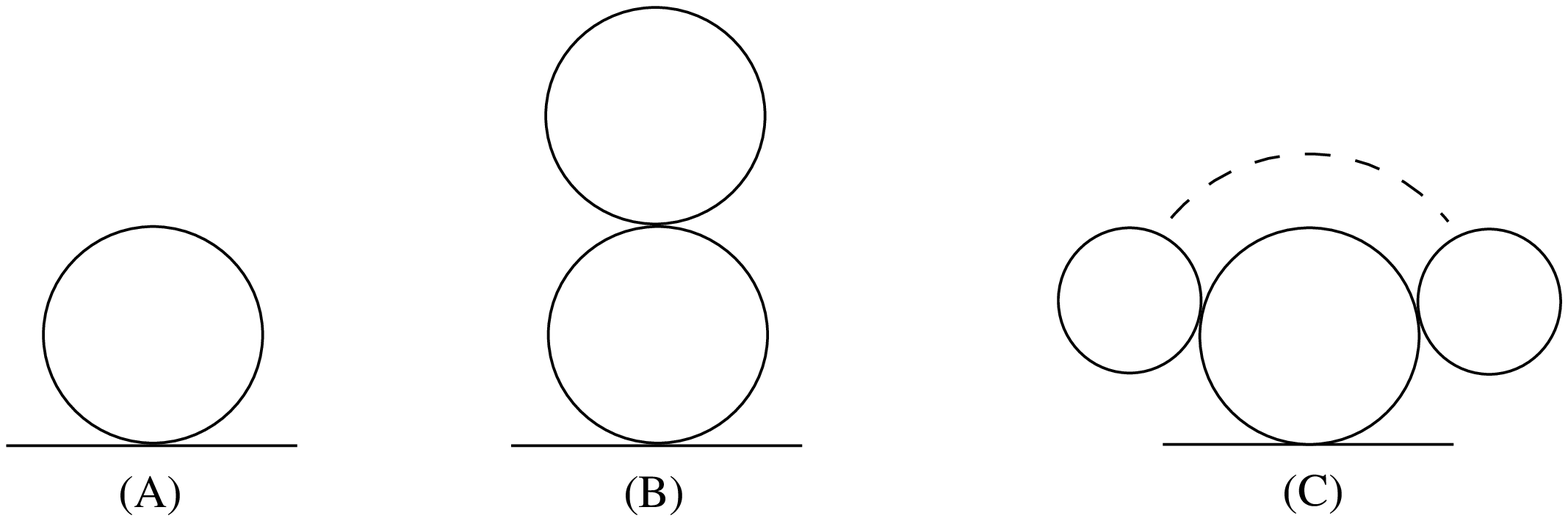}
}
     \caption{Tadpole and cactus diagrams in $\lambda \phi^4$-theory.
	In this theory, only a tadpole diagram become hard thermal loop.}
\label{fig:bubble}
\end{figure}

 In $\lambda \phi^4$ theory at high $T$, only the tadpole diagram 
 in Fig.\ref{fig:bubble} (A) is the hard thermal loop (HTL).
 Therefore, we must resum the cactus-type diagrams in Fig.\ref{fig:bubble}.
 Since they do not depend on external momentum, only the mass term 
 is modified. Thus, the HTLs resummation is performed by shifting the mass 
 term. This happens only when one  considers a theory such as 
 $\lambda \phi^4$ model. If we consider QCD, all vertices with $N$ 
 external gluons and vertices with $N-2$ external gluons 
 and 2 external quarks identified HTLs which must be summed up. 

 In the following subsections (\ref{sec:PMSc} and \ref{sec:FACc}), 
 we adopt the optimization only for the mass term
 in $\lambda \phi^4$ theory for simplicity. 
 We consider the case $\mu^2 < 0$, and the restoration of the symmetry 
 is discussed under various conditions (PMS and FAC, see \S \ref{sec:OPT}).
 The full OPT case (optimization of the mass and coupling terms) is
 discussed in \S \ref{sec:phi4summ}.

\subsection{The PMS condition} 
\label{sec:PMSc}

 Here, we investigate the principle of minimal sensitivity (PMS)
 in one-loop and two-loop orders.
 Since we study the static nature of the phase transition, 
 the thermal effective potential $V(\vphi,m^2)$ is chosen as the relevant 
 physical quantity $O_L$ in Step 3 of \S \ref{sec:OPT}. 

\subsubsection{1-loop analysis}
\label{sec:ph41lPMS}

 For the optimal condition in Step 3 of \S \ref{sec:OPT}, 
 we adopt the following condition at the $O(\delta)$ level:
\beq
\label{eq:V1PMS}
	{\d V^{0 + \delta}(\vphi,m^2) \over \d m^2} =0,
\eeq
 where $V^{0 + \delta}(\vphi,m^2) = V^0 + V^{\delta}$
 (see Eq.(\ref{eq:fullV}) in Appendix \ref{app:fullV}). 
 However, this condition does not lead to the appropriate gap equation.

 Differentiation with respect to $m^2$ corresponds to cutting one of 
 the internal lines of $V(\vphi,m^2)$. This is because the power of 
 the propagator is raised by 1.
 As one can easily see from Fig.\ref{fig:V1} in Appendix \ref{app:2-loop}, 
 cutting the internal line of 
 $V(\vphi,m^2)$ cannot produce HTLs (Fig.\ref{fig:bubble} (A)). 
 Therefore, Eq.(\ref{eq:V1PMS}) cannot sum the tadpole type diagrams, and
 it is not meaningful to adopt the PMS condition at the $O(\delta)$ level.
 Thus, we need to go to the next order, which is two loops.

\subsubsection{2-loop analysis}
\label{sec:phi42P}

 The PMS condition for the 2-loop effective potential reads
\beq
\label{eq:V2PMS}
	{\d V^{0 + \delta + \delta^2}(\vphi,m^2) \over \d m^2} =0,
\eeq
 where $V^{0 + \delta + \delta^2}(\vphi,m^2) = 
 V^0 + V^{\delta} + V^{\delta^2}$. Cutting one of the internal lines of 
 Fig.\ref{fig:V2} (a) leads to HTLs in the scalar theory. 
 The explicit form of (\ref{eq:V2PMS}) is given in Appendix \ref{app:PMS}.
 At high $T$ (in the symmetric phase), Eq.(\ref{eq:V2PMS}) reduces to 
\beq
\label{eq:V2PMSre}
	\left. {\d V^{0 + \delta + \delta^2}(\vphi,m^2) \over \d m^2}
	\right|_{\vphi = 0} 
	\stackrel{\beta m^2 \rightarrow 0}{\longrightarrow}
	(\chi - {\lambda T^2 \over 24}) {T \over 16 \pi m}.
\eeq 
 This equation gives the solution
\beq
\label{eq:sV2PMSre}
	m^2(T) = \mu^2 + {\lambda T^2 \over 24}.
\eeq
 Since Eq.(\ref{eq:sV2PMSre}) corresponds to the Debye screening mass 
 at high $T$, the condition (\ref{eq:V2PMS}) correctly resums higher 
 order terms and recovers the reliability of the perturbation theory 
 at finite $T$.

 To determine the vacuum, we must also solve the equation
\beq
\label{eq:statV}
	\left. {\d V(\vphi,m^2) \over \d \vphi} 
	\right|_{\vphi = \vphi_0} = 0.
\eeq
 The derivative with respect to $\vphi$ does not act on $m^2$
 by definition, even if $m^2(\vphi)$ depends on $\vphi$.
 However, the gap equation Eq.(\ref{eq:V2PMS}) leads to
\beq
\label{eq:phi4Dd}
	{d V(\vphi,m^2(\vphi)) \over d \vphi} & = & 
	{\d V(\vphi,m^2) \over \d \vphi} 
	+ {\d V(\vphi,m^2) \over \d m^2} {\d m^2 \over \d \vphi}
	\nonumber \\
	& = & {\d V(\vphi,m^2) \over \d \vphi}.
\eeq
 Thus, in this case, the total derivative with respect to $\vphi$ is
 equal to the partial derivative. 
 This is an advantage of the PMS condition over the FAC condition 
 in studying the phase transition if $m^2(\vphi)$, which is the solution 
 of Eq.(\ref{eq:V2PMS}), has physical meaning for all $\vphi$.

\vspace*{0.2cm}
\noindent
{\it Initial condition}
\vspace*{0.2cm}

 We solved Eqs.(\ref{eq:V2PMS}) and (\ref{eq:statV}) numerically.
 There are three parameters in these equations, 
 $\mu^2$, $\lambda$ and $\kappa$.
 ($m^2$ is determined by Eq.(\ref{eq:V2PMS}).)
 Since we assume that the loop expansion at $T=0$ is a valid approximation,
 the renormalization point $\kappa$ is chosen so that $\mu^2 = m^2$
 is satisfied. Thus, the OPT is equivalent to the naive-loop expansion 
 at $T=0$.
 In other words, we use the OPT only for resummation at finite $T$ to
 recover the reliability of the perturbation theory at finite $T$.
 (Note, however, that the relation $\mu^2 = m^2$ at $T=0$ must be satisfied
 in Eq.(\ref{eq:V2PMS}).)
 Although the explicit values of the parameters are not important for 
 our qualitative study (in fact, these parameters are normalized 
 by their initial values), the initial values, 
 $\lambda = 10.0$ and $\vphi_0 = 10.0$ were used for simplicity.
 As a result of solving Eqs.(\ref{eq:V2PMS}) and (\ref{eq:statV}) 
 simultaneously, we obtain $\mu^2 = m^2 = -170$, $\kappa^2 = 87.6$ and 
 $M^2 = m^2 + {\lambda \over 2} \vphi_0^2 = 330$.

\vspace*{0.2cm}
\noindent
{\it Results of numerical calculation}
\vspace*{0.2cm}

\begin{figure}[ht]
\centerline{
    \epsfysize=5.5cm 
    \epsfbox{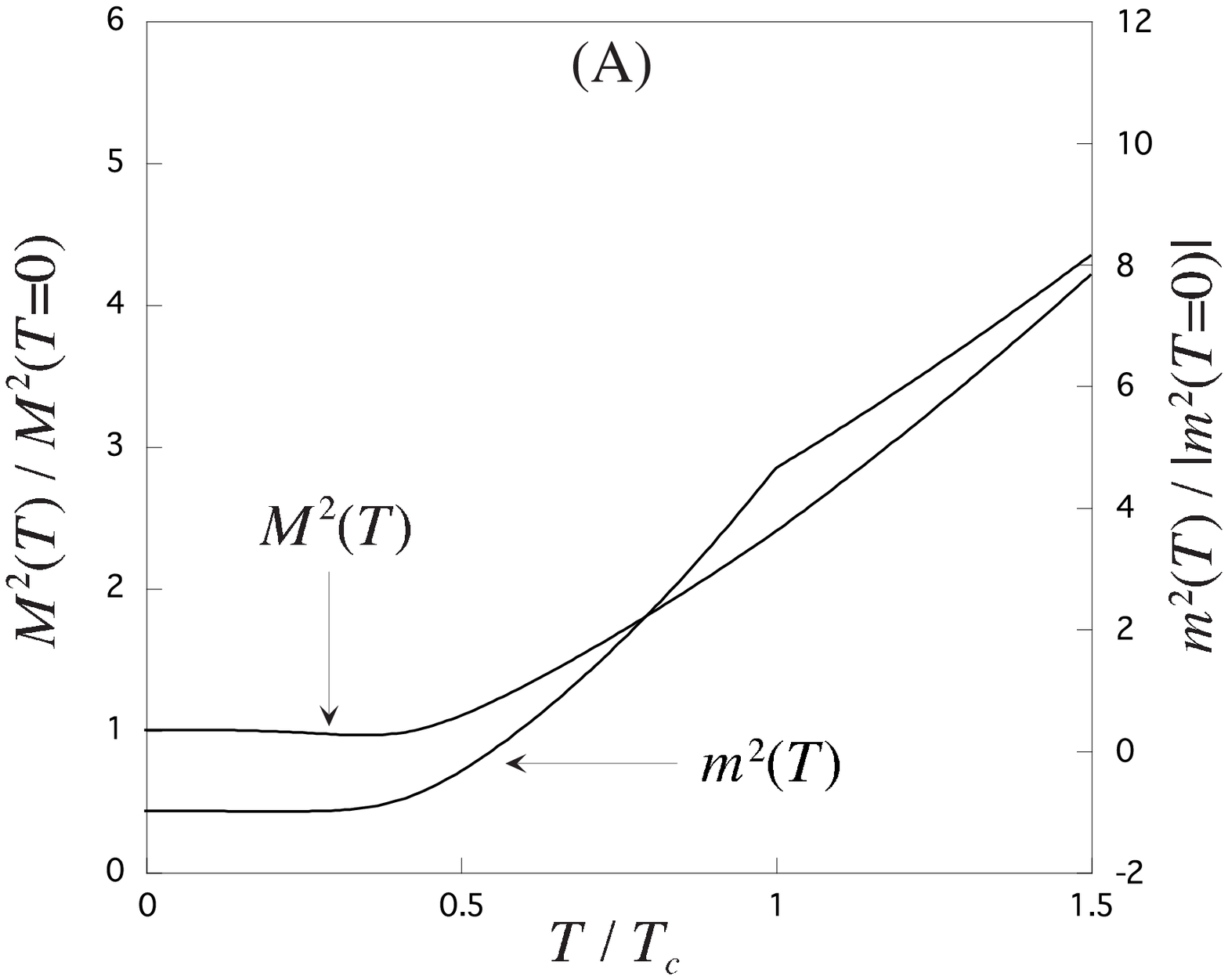} \
    \epsfysize=5.5cm 
    \epsfbox{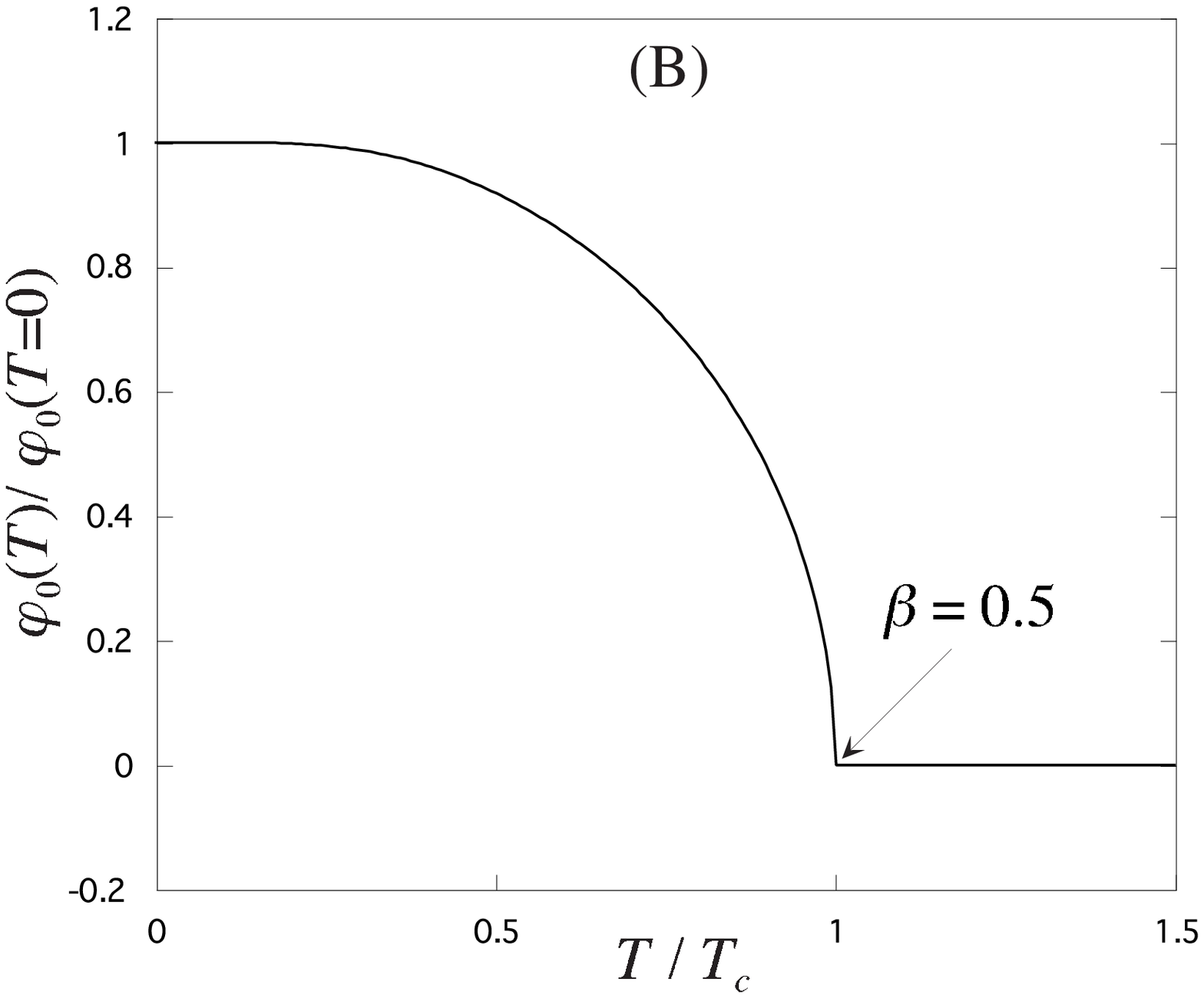}
}
     \caption{(A) The tree level mass $M^2 (T)= m^2 + \lambda \vphi_0 /2$, 
	with the left vertical scale and the mass parameter $m^2(T)$, 
	with the right vertical scale obtained with the PMS condition.
	They are normalized by their vales at $T=0$.
	(B) Vacuum expectation value $\vphi_0$ normalized
	by $\vphi_0(T=0)$.}
\label{fig:pms-main}
\end{figure}

 The results of the numerical calculations with the PMS condition are shown 
 in Fig.\ref{fig:pms-main}. 
 Figure \ref{fig:pms-main} (A) plots the tree-level mass 
 $M^2(T) = m^2(T) + \lambda \vphi_0 /2$ with the left vertical scale 
 and the optimized parameter $m^2(T)$ with the right vertical scale.
 $M^2(T)$ is clearly not tachyonic for any $T$. This result,
 Eq.(\ref{eq:sV2PMSre}), confirms that OPT with the two-loop PMS condition 
 for $V(\vphi, m^2)$ is successful for the resummation of HTLs.

\begin{figure}[ht]
\centerline{
    \epsfysize=3.8cm
    \epsfbox{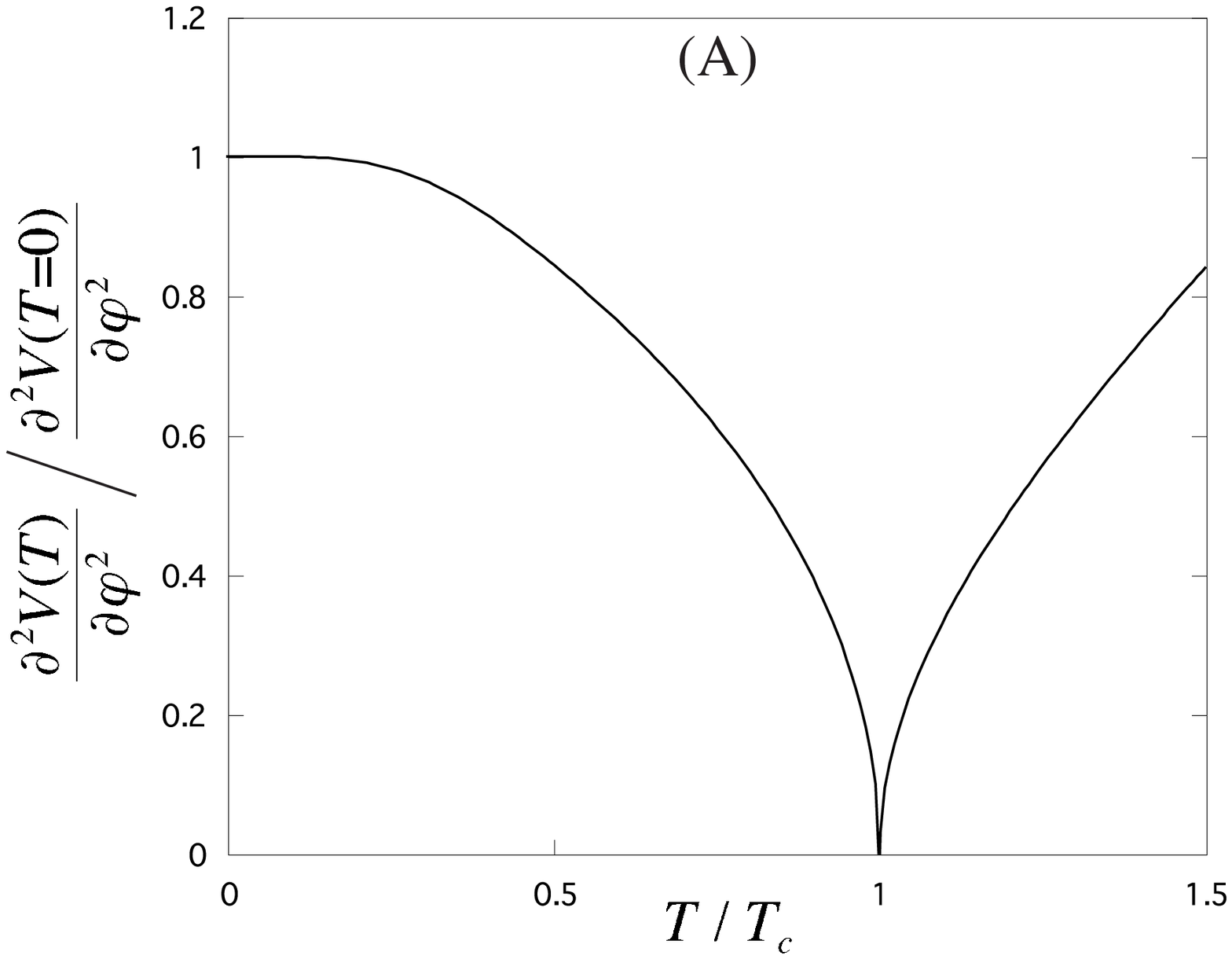} \ 
    \epsfysize=3.8cm 
    \epsfbox{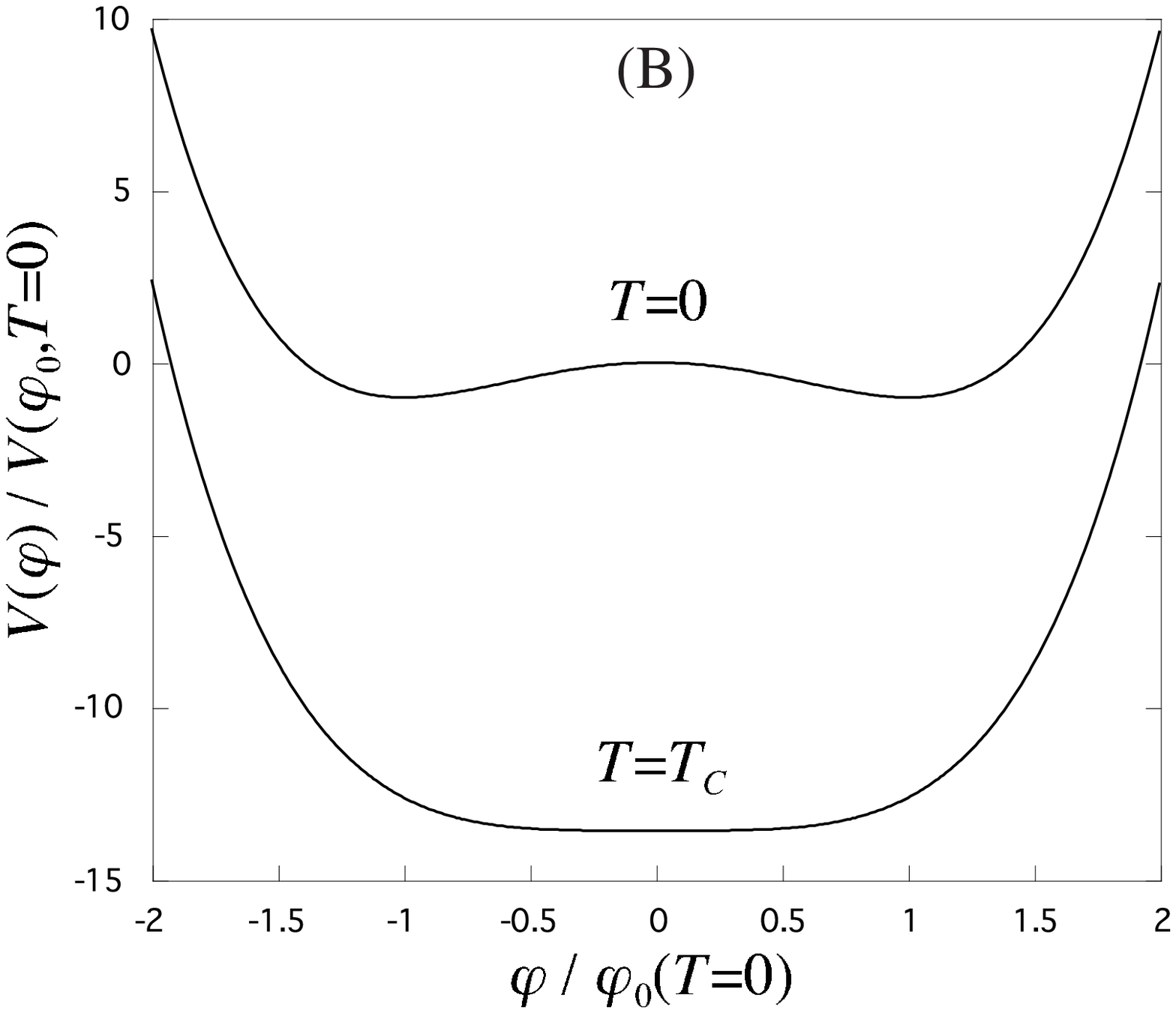} \ 
    \epsfysize=3.8cm 
    \epsfbox{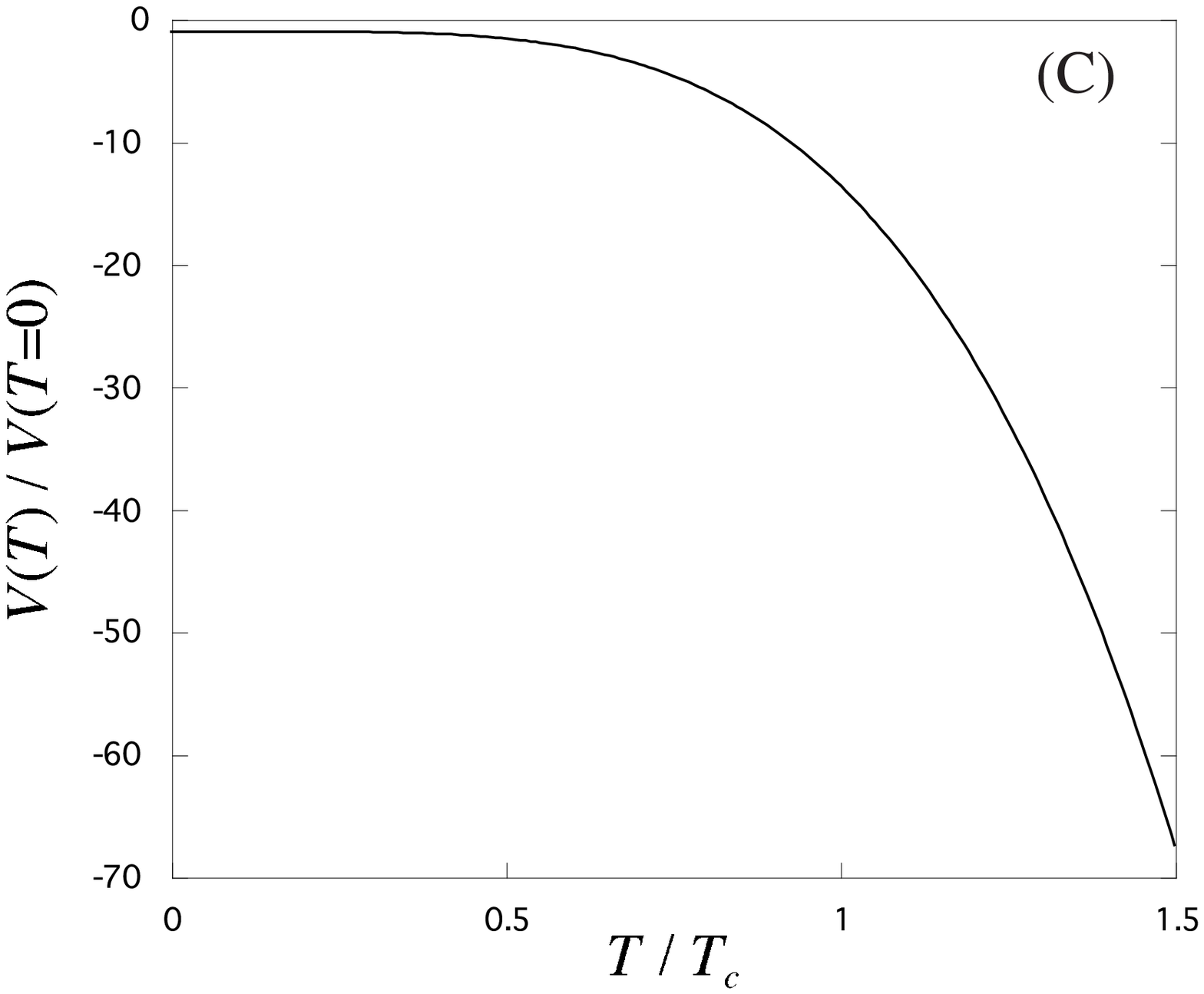}
}
     \caption{(A) Second derivative of $V(T)$ with respect to $\vphi$ 
	for the PMS condition. 
	(B) Effective potentials at $T=0$ and $T=T_c$.
	(C) Minimum value of the effective potential as a function of $T$.}
\label{fig:pms-sub}
\end{figure}

 Figure \ref{fig:pms-main} (B) shows the $T$ dependence of 
 the thermal expectation value $\vphi_0 (T)$
 divided by $\vphi_0$ at $T = 0$. From this result,
 the phase transition is found to be of second order.
 Figure \ref{fig:pms-sub} (A) shows the second derivative of $V(\vphi, m^2)$ 
 with respect to $\vphi$ at $\vphi = \vphi_0(T)$ as a function of $T/T_c$.
 This figure also shows the second order nature of the transition.
 When the transition is of second order, the effective potential becomes 
 flat for $\vphi_0(T) = 0$ at the critical temperature $T_c$, 
 which is expressed by the following equation:
\beq
\label{eq:d2V=0}
	\left. {\d^2 V(\vphi, m^2) \over \d \vphi^2} \right|_{\vphi =0} = 0.
\eeq
 The effective potentials at $T=0$ and $T_c$ are shown 
 in Fig.\ref{fig:pms-sub} (B). We can also confirm the second-order nature
 of the phase transition from Fig.\ref{fig:pms-sub} (B).
 We note here that a second order transition for the $\lambda \phi^4$-theory 
 is expected from the renormalization group analysis \cite{renormGF} and 
 lattice simulations at finite $T$. \cite{latticephi4}

 We found that the critical exponent $\beta$, which is defined by
\beq
\label{eq:defbeta}
	\vphi_0 (T) \propto \left| { T - T_c \over T_c} \right|^{\beta},
\eeq
 becomes $0.5$ in the two-loop analysis. This is the value expected 
 from the Landau mean-field theory, which implies that our approximation 
 is still within the level of the mean-field theory.
 Since the OPT corresponds to a generalized mean field theory, this was
 in fact anticipated. 

 In Fig.\ref{fig:pms-sub} (C), the minimum value of the thermal effective 
 potential $V(\vphi_0, m^2)$, which is equivalent to the Gibbs free energy,
 is shown. Its value decreases monotonically as $T$ increases. 


\subsection{The FAC condition}
\label{sec:FACc}

 In this subsection, we apply the FAC condition. 
 The simplest condition to resum the HTLs in this case is
\beq
\label{eq:FAC}
	\Sigma_R(\omega =0, |\vec{k}|=0;T) = 0,
\eeq
 where $\Sigma_R$ is a retarded two-point self-energy which is defined as
\beq
\label{eq:defRS}
	\Sigma_R(\omega, \vec{k};T) = 
	\left. {\d^2 V(\vphi, m^2) \over \d \vphi^2} \right|_{\vphi = \vphi_0}.
\eeq
 In the following, we investigate
 the above condition at one-loop and two-loop orders.

\subsubsection{1-loop analysis}

 The HTLs can be resummed even with the one-loop FAC condition of 
 Eq.(\ref{eq:FAC}).
 For this reason, this condition has been adopted in many studies
 for simplicity, and it allows the tachyon problem to be solved at low $T$. 
 However, the phase transition becomes first order with this condition.
 Here, we do not recapitulate these results, because they are 
 discussed extensively in the literature. Instead, we examine 
 the two-loop condition Eq.(\ref{eq:FAC}) in \S \ref{sec:phi42lF} 
 to compare the result with 
 that for the PMS condition discussed in the \S \ref{sec:phi42P}.

\subsubsection{2-loop analysis}
\label{sec:phi42lF}

 Since the physics should not depend on the artificial parameter $m^2$,
 one expects that the result should not depend on the choice of the optimized 
 conditions. However, the 2-loop PMS condition Eq.(\ref{eq:V2PMS}) 
 leads to a second-order phase transition, and 1-loop the FAC condition is 
 known to lead to a first-order transition. 
 Therefore, it is necessary to study whether the FAC condition at the 
 two-loop level gives a second order phase transition.

 For the FAC condition at $O(\delta^2)$, we adopt
\beq
\label{eq:V2FAC}
	\Sigma_R^{\delta^2}(\omega, \vec{k};T) = 
	\left. {\d^2 V^{\delta^2} (\vphi, m^2) \over \d \vphi^2} 
	\right|_{\vphi = \vphi_0} = 0.
\eeq
 An explicit formula for $V^{\delta^2}$ is given in Appendix.\ref{app:fullV}.
 At high $T$, Eq.(\ref{eq:V2FAC}) is reduced to
\beq
\label{eq:V2FACh}
	\left. {\d^2 V^{\delta^2} \over \d \vphi^2} \right|_{\vphi = 0}
	& \stackrel{\beta m^2 \rightarrow 0}{\longrightarrow} & 
	{\lambda T \over 16 \pi m} (\chi - {\lambda T^2 \over 24})
	+ {\lambda^2 m^2 \over 2(4 \pi)^4} \{ c + (-2 + {1 \over 2}
	\ln{m^2 \over \kappa^2}) \ln{m^2 \over \kappa^2} \} \nonumber \\
	&& + {\lambda^2 T^2 \over 24 (4 \pi)^2} 
	(3.30 -\ln{T^2 \over \kappa^2}).
\eeq
 The first term on the right hand side (r.h.s.) produces the resummation 
 of the HTLs.

\vspace*{0.2cm}
\noindent
{\it Initial conditions}
\vspace*{0.2cm}

 The initial parameters (at $T=0$) are chosen as $\mu^2 = m^2$, 
 $\lambda = 10.0$ and $\vphi_0 = 10.0$ as in the previous section. 
 The condition $\mu^2 = m^2$ is needed for agreement with the naive 
 loop-expansion at $T=0$. Note that we recognize that the loop expansion 
 at $T=0$ is a valid approximation. The other resultant values, 
 which are determined by solving Eqs.(\ref{eq:statV}) and 
 (\ref{eq:V2FAC}) simultaneously, 
 are $\mu^2 = m^2 = -166$, $\kappa^2 = 137$ and 
 $M^2 = m^2 + {\lambda \over 2} \vphi_0^2 = 334$.

\vspace*{0.2cm}
\noindent
{\it Results of the numerical calculation}
\vspace*{0.2cm}

\begin{figure}[ht]
\centerline{
    \epsfysize=5.5cm 
    \epsfbox{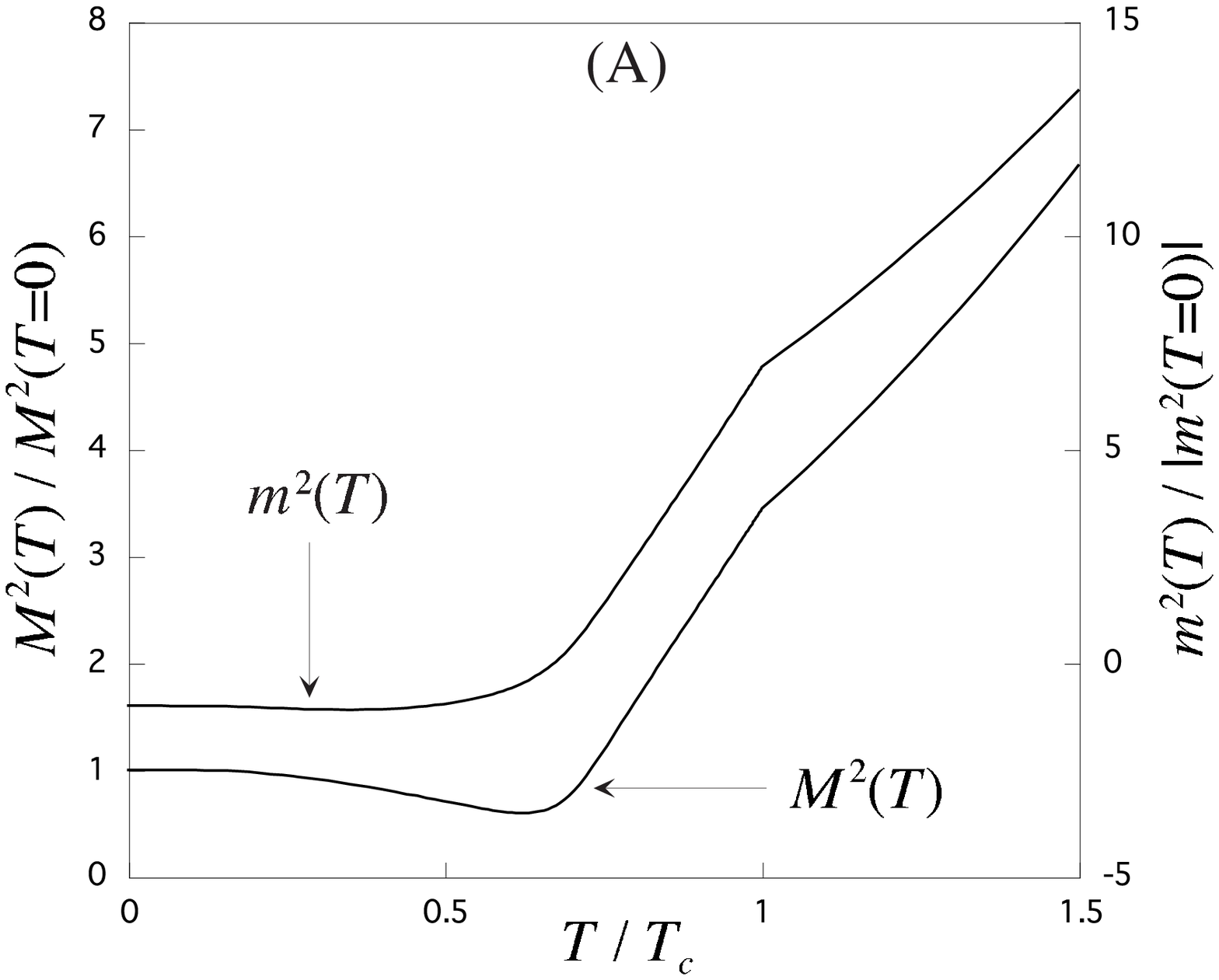} \ 
    \epsfysize=5.5cm 
    \epsfbox{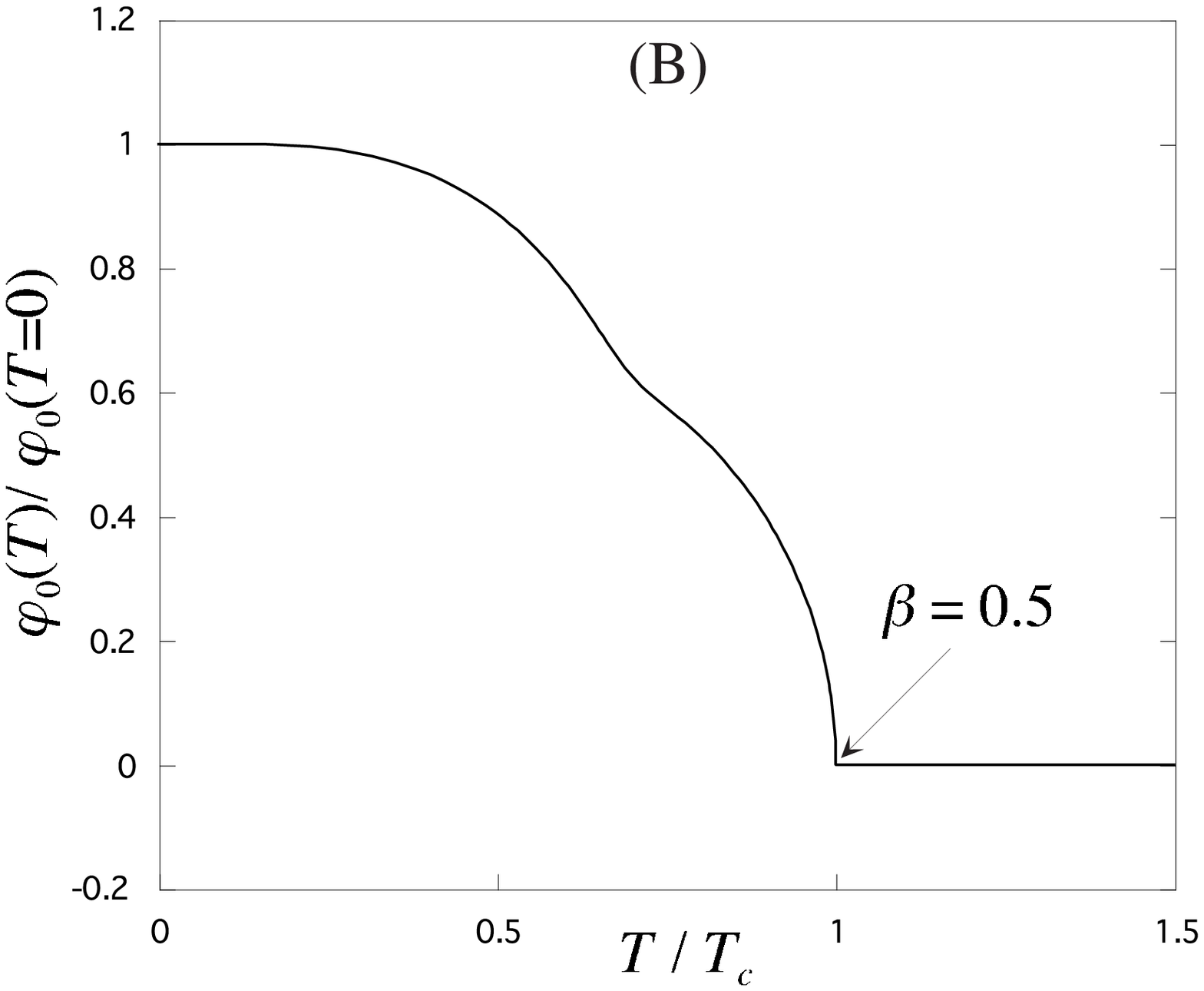}
}
     \caption{(A) The tree level mass $M^2 (T)= m^2 + \lambda \vphi_0 /2$, 
	with the left vertical scale, and the mass parameter $m^2(T)$, with 
	the right vertical scale, in the case of the FAC condition.
	They are normalized by their values at $T=0$.
	(B) Vacuum expectation value $\vphi_0$ normalized by
	$\vphi_0(T=0)$. }
\label{fig:fac-main}
\end{figure}

\begin{figure}[ht]
\centerline{
    \epsfysize=3.8cm
    \epsfbox{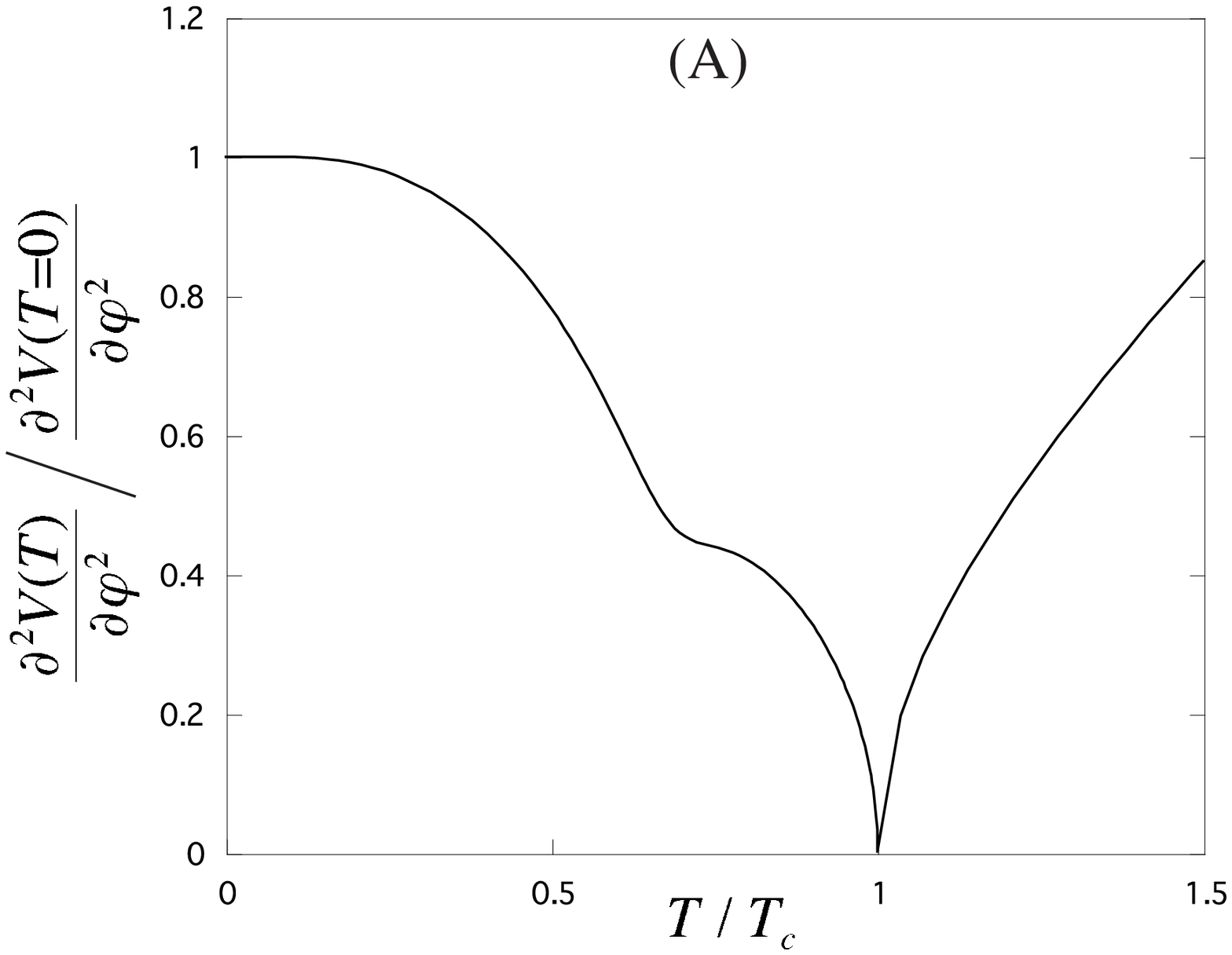} \
    \epsfysize=3.8cm 
    \epsfbox{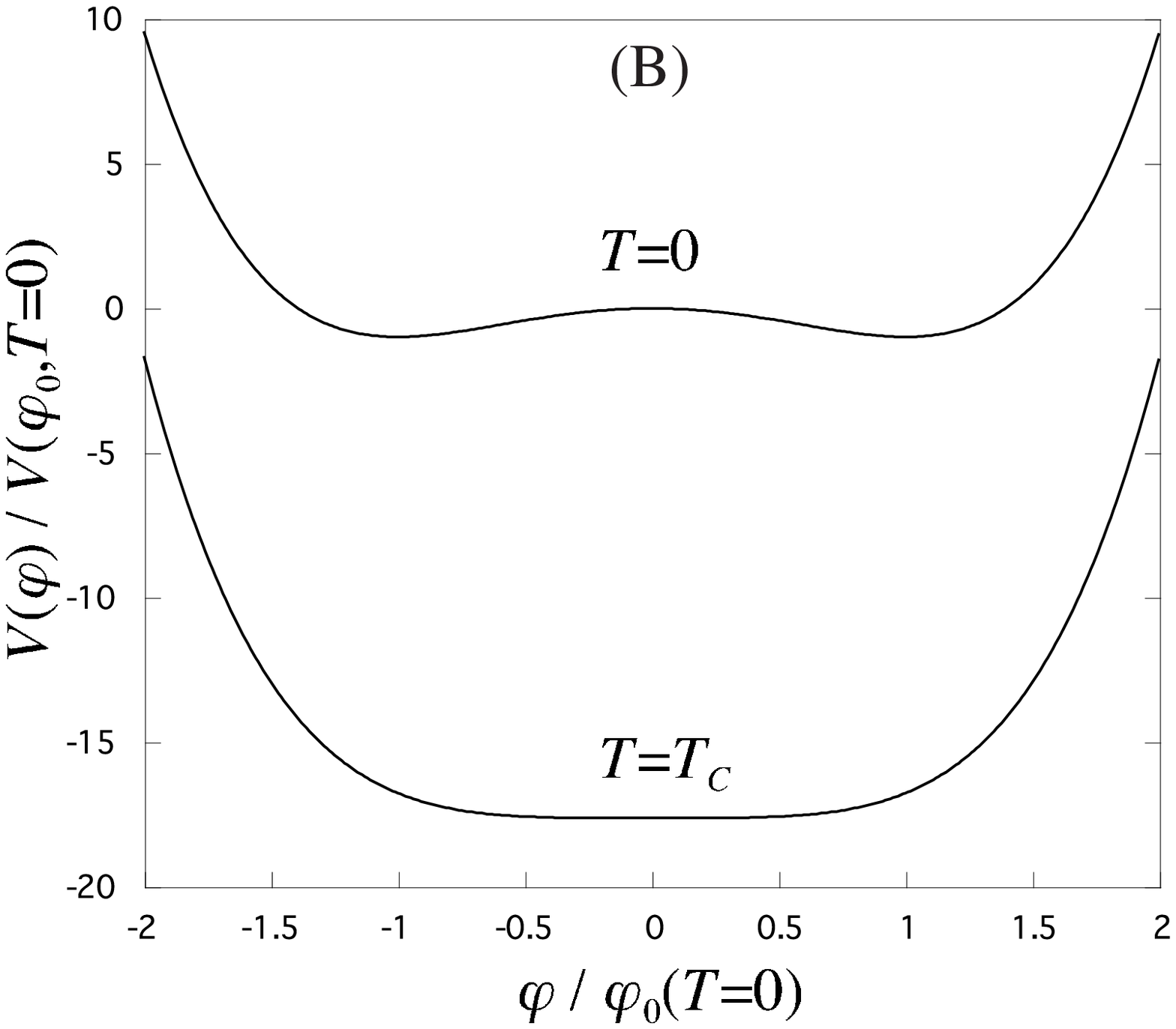} \ 
    \epsfysize=3.8cm
    \epsfbox{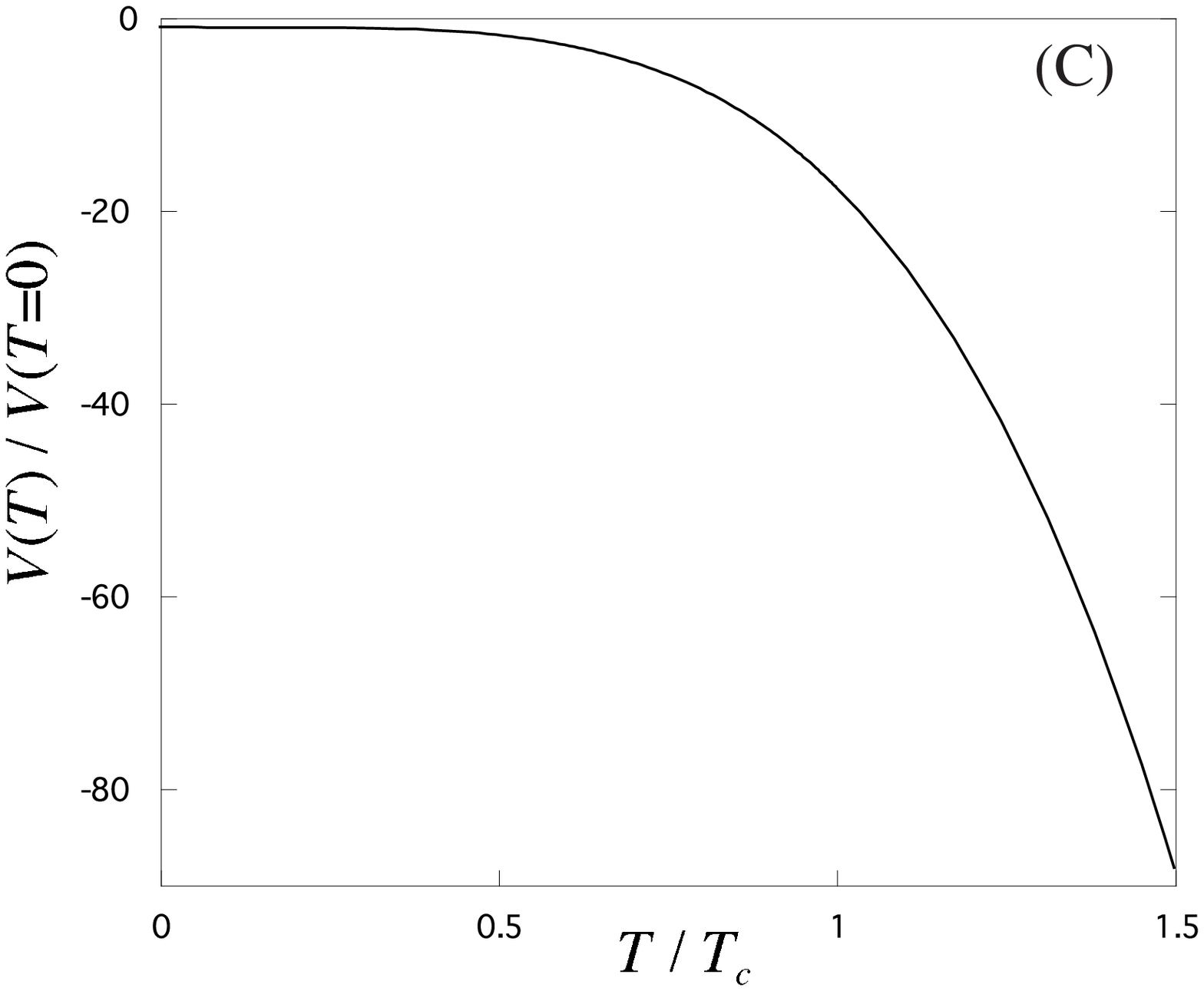}
}
     \caption{(A) Second derivative of $V(T)$ with respect to $\vphi$ 
	for the FAC condition. 
	(B) Effective potential at $T=0$ and $T=T_c$.
	(C) Minimum value of the effective potential as a function of $T$.}
\label{fig:fac-sub}
\end{figure}

 The results are given in Figs.\ref{fig:fac-main} and \ref{fig:fac-sub}.
 One can see that the qualitative features of the figures are the same as
 the PMS  results in Figs.\ref{fig:pms-main} and \ref{fig:pms-sub} 
 \footnote{The physical reason for the shoulder structure around 
 $T/T_c \simeq 0.7$ in Figs.\ref{fig:fac-main} (B) and \ref{fig:fac-sub}
 (A) is not yet understood.}.
 From Fig.\ref{fig:fac-main} (A), one can see that the tachyon problem 
 is also avoided in this case. Figures \ref{fig:fac-main} (B),
 \ref{fig:fac-sub} (A) and  \ref{fig:fac-sub} (B) show the second 
 order phase transition with $\beta = 0.5$. In this case,
 the Gibbs free energy decreases uniformly (Fig.\ref{fig:fac-sub} (C)). 
 Thus, the 2-loop condition in both PMS and FAC give qualitatively 
 the same results. This is the desired property, and it
 shows the validity of the OPT.


\subsection{Some remarks}
\label{sec:phi4summ}

 We tested the two conditions represented by Eqs.(\ref{eq:V2PMS}) 
 and (\ref{eq:V2FAC}) 
 in the OPT to study the phase transition for $\lambda \phi^4$ theory. 
 The two conditions give qualitatively the same results and show that 
 the resummations were successfully performed.
 However, there are three remarks in order, which we give below.

\vspace*{0.2cm}
\noindent
{\it FAC with $O(\delta) + O(\delta^2)$}
\vspace*{0.2cm}

 One may try to use the two-loop FAC condition
\beq
\label{eq:V2FACl}
	\Sigma_R^{\delta + \delta^2}(\omega, \vec{k};T) = 
	\left. {\d^2 V^{\delta + \delta^2} (\vphi, m^2) \over \d \vphi^2} 
	\right|_{\vphi = \vphi_0} = 0
\eeq
 instead of Eq.(\ref{eq:V2FAC}). This condition implies that all the loop 
 corrections vanish at zero external momentum. 
 In other words, $m^2$ contains all contributions up to two-loop order. 
 From Eq.(\ref{eq:V2FACl}), the following equation is obtained:
\beq
\label{eq:V2FAClr}
	{\d^2 V^{0 + \delta + \delta^2} \over \d \vphi^2} 
	= M^2 = m^2 + {\lambda \over 2} \vphi^2.
\eeq
 At $T = T_c$, if one assumes a second-order phase transition,
 Eq.(\ref{eq:V2FAClr}) must be zero; that is $m^2 (= M^2)= 0$.
 However, the vanishing of the tree-level mass causes an infrared 
 divergence in the loop
 integrals. Actually, the left-hand side (l.h.s) of Eq.(\ref{eq:statV}) 
 diverges as $M^2 \rightarrow 0$. 
 Thus, $M^2=0$ is never satisfied at $T=T_c$. 
 From this argument, we see that a second-order phase transition 
 cannot be realized with the condition given by Eq.(\ref{eq:V2FAClr}).
 \footnote{In Ref.\cite{BM}, Eq.(\ref{eq:V2FACl}) is used as the FAC 
 condition. Thus, the calculation there leads to a first-order phase 
 transition. (See also, Ref.\cite{mallikmiss}.)}

\vspace*{0.2cm}
\noindent
{\it Full OPT}
\vspace*{0.2cm}

 The full OPT, which includes an optimization of the coupling constant,
 allows the possibility not only of avoiding the above infrared problem 
 but also of going beyond the mean field approximation near $T_c$. 
 Therefore, it is very interesting to explore it.
 However, the PMS condition for $V(\vphi, m^2, g)$
 requires tree-loop calculation. This can be understood as follows.
 Suppose one chooses the PMS condition at the two-loop level as
\beq
\label{eq:e-fullPMS}
	{\d V^{0 + \delta + \delta^2}(\vphi,m^2,g) \over \d m^2} = 0, \ \ 
	{\d V^{0 + \delta + \delta^2}(\vphi,m^2,g) \over \d g} = 0.
\eeq
 In the symmetric phase ($\vphi=0$), Eq.(\ref{eq:e-fullPMS}) can be reduced to
\beq
\label{eq:fPMSm}
	\left. {\d V \over \d m^2} \right|_{\vphi =0} & = &
	{\d K_t \over \d M^2} ( \chi + g K_t), \\
\label{eq:fPMSg}
	\left. {\d V \over \d g} \right|_{\vphi =0} & = &
	{1 \over 2} K_t^2.
\eeq
 From Eq.(\ref{eq:fPMSg}), $K_t = 0$ is obtained. This gives a solution
 for $m^2$. (Note that $K_t$ depends on $g$ only through
 $M^2 = m^2 + g \vphi^2 /2$.)
  Then, this solution is substituted into Eq.(\ref{eq:fPMSm}), giving
\beq
\label{eq:fPMSmr}
	\left. {\d V(\vphi,m^2,g) \over \d m^2} \right|_{\vphi =0} = 
	{\d K_t \over \d M^2}  \chi = 0.
\eeq
 Since this relation also does not depend on $g$, we cannot determine 
 $g$. Thus, Eq.(\ref{eq:e-fullPMS}) cannot give a solution for
 $m^2$ and $g$ simultaneously in the symmetric phase. 

 On the other hand, in the three-loop calculation, 
 $V(\vphi, m^2, g)$ has the term
\beq
\label{eq:V3t3}
	- g {\d K_t \over \d M^2} (\chi + {g \over 2} K_t).
\eeq
 This leads to the HTLs (tadpole diagram) in $\d V / \d g$,
 and depends on $g$ even in the symmetric phase.
 Thus, the full OPT with the PMS condition for the tree-loop 
 effective potential 
 has the possibility of giving a solution for $m^2$ and $g$ simultaneously.

 What about the FAC condition in full OPT?
 Suppose we use the following FAC condition:
\beq
\label{eq:fullFAC}
	{\d^2 V^{\delta^2}(\vphi,m^2,g) \over \d \vphi^2} = 0, \ \ 
	{\d^4 V^{\delta^2}(\vphi,m^2,g) \over \d \vphi^4} = 0.
\eeq
 Unfortunately, we could not find a solution which improves the previous 
 results near $T_c$ at the two-loop level. 
 We have tried all possible variations
 ($V^{\delta^2}(\vphi,m^2,g)$ in Eq.(\ref{eq:fullFAC})
 replaced by $V^{\delta + \delta^2}(\vphi,m^2,g)$, 
 $V^{\delta}(\vphi,m^2,g)$, and so on), but
 improved solutions were not obtained for the critical exponent.
 Thus, two-loop order, it seems that it is not possible to improve
 the previous analysis near $T = T_c$ in the full OPT with the FAC condition. 
 In any case, further study along this line is necessary.

\vspace*{0.2cm}
\noindent
{\it Limiting temperature}
\vspace*{0.2cm}

 For sufficiently high $T$, there are no solutions with
 (\ref{eq:V2PMS}) and (\ref{eq:V2FAC}) for $m^2(T)$, 
 because logarithmic terms of the form $\ln (T / \kappa)$ dominate. 
 This means that the renormalization point $\kappa$, which is fixed at $T=0$,
 becomes a bad choice as $T$ increases.
 To avoid this situation, one may try a renormalization group improvement. 
 Since the typical ``scale'' of this system is $T$,
 one may choose $\kappa = T$. In this case,
 no large $\ln (T / \kappa)$ appears.
 This renormalization group improvement would extend the applicability of OPT. 
 However, in non-asymptotically free theories, such as $\lambda \phi^4$ and
 $O(N)$ $\lambda \phi^4$ theories, there eventually appears a Landau pole
 where the running coupling constant $\lambda (\kappa = T)$
 diverges at a certain $T$. \cite{triv,latticetri}
 Thus, the theory is not applicable beyond this $T$.

\section{Summary}
\label{sec:summary}

 In this paper, we have generalized the optimized perturbation theory 
 (OPT) at finite temperature ($T$) in the $\lambda \phi^4$ theory to 
 incorporate the optimization of the coupling constant. 
 The naive loop expansion for theories with spontaneous symmetry breaking
 is known to break down at low $T$ (tachyon pole problem) and 
 high $T$ (hard thermal loops (HTLs) problem).
 By contrast, we have shown that the OPT does not suffer from such
 difficulties. This is because that the OPT resums the higher-order terms 
 of the naive perturbation theory in a consistent way by imposing appropriate 
 conditions, such as the principle of minimal sensitivity (PMS) and
 the fastest apparent convergence (FAC).

 The advantages of the OPT over other self-consistent methods are that 
 one can carry out the renormalization of ultraviolet divergences 
 systematically and that the Nambu-Goldstone (NG) theorem at finite $T$ is 
 trivially satisfied at any given order.

 In \S \ref{sec:phase}, we applied the OPT to $\lambda \phi^4$ theory
 to examine whether it can describe the finite $T$ phase transition 
 correctly. Carrying out the two-loop computation of the effective
 potential in the OPT, we have found that both PMS and FAC conditions give 
 the correct second order transition. The critical exponent $\beta$,
 however, is found to coincide with the mean-field value at this level. 
 The full OPT, where both the mass and the coupling constant are 
 shifted, may or may not improve the result. This remains as an open problem.

 Application of the idea of the OPT to gauge theories at finite $T$ is
 also an interesting problem. However, to go beyond the hard thermal 
 loops (HTLs) resummation scheme, one must solve the following two problems.
\begin{itemize}
\item Since an infinite number of $N$-point vertex functions
	($N$-gauge boson vertices and $(N-2)$-gauge boson 2-fermion vertices)
	identified with HTLs in gauge theories and a naive mass term breaks 
	the gauge symmetry, one must take into account an infinite number of 
	effective vertices like HTLs resummation scheme. 
	However, this infinite number vertices 
	may create a difficulty for the renormalization.
\item It is known that HTLs resummation scheme breaks down at high $T$ 
	due to the so-called magnetic mass problem. \cite{linde80}
\end{itemize}
 Therefore, we may need further generalization of the OPT to apply it for
 gauge theories.

\section*{Acknowledgements}

 I am greatly indebted to T. Hatsuda for his helpful discussions
 and encouragement. I also wish to thank all the members of the theoretical 
 nuclear physics group at the University of Tsukuba, the Yukawa Institute 
 for Theoretical Physics, and the nuclear theory group at Kyoto University.
 I would also like to thank the Japan Society of Promotion for the 
 Science (JSPS) for financial support.

\appendix
\section{2-Loop Effective Potential in $\lambda \phi^4$-Theory}
\label{app:2-loop}

 Here, we calculate the physical effective potential in (\ref{eq:ala1})
 in the real-time formalism, \cite{real} which is equivalent to 
 the effective potential defined in the imaginary-time formalism.  
 Following Ref. \cite{realeff}, we compute the tadpole functions 
 $V^{(1,0)}$, \footnote{The ``$(1,0)$'' in $V^{(1,0)}$ indicates 
 the $\vphi_1$ tadpole function. $\vphi_1$ is a type 1 field in the 
 real-time formalism.}
 and then the effective potential is integrated over
 the classical field $\vphi$.\cite{Wtado}
 Since the propagator in the $T=0$ part and $T \neq 0$ 
 part decouples in the real time formalism, calculations of the $T=0$ parts 
 yield the same results as the ordinary $T=0$ perturbation theory. 
 Hence, we can calculate the $T$-dependent term and $T$-independent 
 term separately.

\begin{figure}[ht]
\centerline{
    \epsfysize=3cm 
    \epsfbox{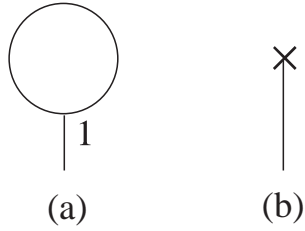}
}
     \caption{(a) One-loop diagram contributing to the 1-point function.
	(b) Counterterm contribution 
 	(which includes all $\hbar$-order contributions in principle).
	The index 1 indicates a type 1 vertex. 
	The cross represents the counterterms.}
\label{fig:1-loop}
\end{figure}

\begin{figure}[ht]
\centerline{
    \epsfysize=4.5cm 
    \epsfbox{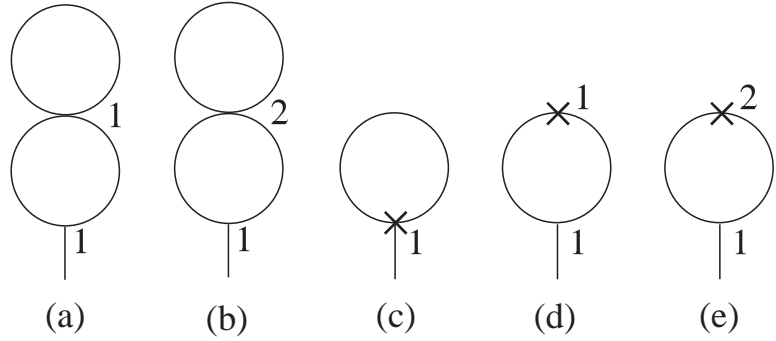}
}
     \caption{(a) and (b) 2-bubble diagrams.
	(c)-(e) Tadpole diagrams including counterterms.}
\label{fig:2tado}
\end{figure}

\begin{figure}[ht]
\centerline{
    \epsfysize=3cm 
    \epsfbox{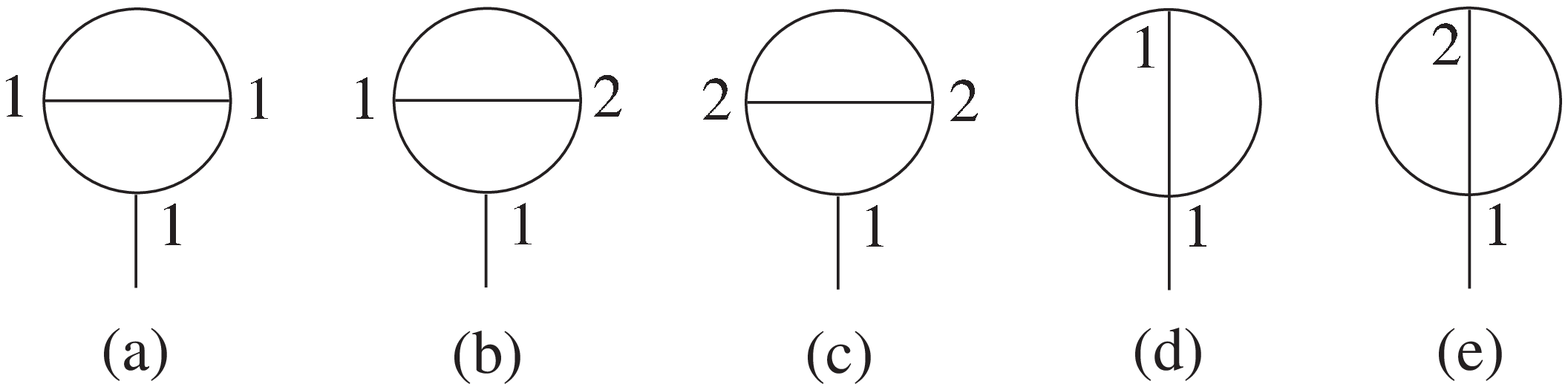}
}
     \caption{Diagrams which contribute to the setting sun diagram.}
\label{fig:setsun}
\end{figure}

 The Feynman diagrams contributing to the $\vphi_1$ tadpole functions 
 up to the two-loop level are shown in 
 Figs.\ref{fig:1-loop}, \ref{fig:2tado}, \ref{fig:setsun}.
 The vertex with the number 1 (2) represents the type 1 (2) field 
 self-interaction. The type 2 field is necessary to cancel 
 pathological pinch singularities. 

 The formulas of the tadpole functions read
\beq
\label{eq:tadoform}
	V^{(1,0)}  & = & V^{(1,0)1} + V^{(1,0)2}, \\
\label{eq:dV1}
	-iV^{(1,0)1} & = & -i {\lambda \vphi_1 \over 2} \kappa^{2 \vep}
	\int {d^n k \over (2 \pi)^n} iD_{\beta}^{11}(k) + counterterms \\
\label{eq:2tadoform}
	V^{(1,0)2} & = & V^{(1,0)2}_{2bubble} + V^{(1,0)2}_c 
			+ V^{(1,0)2}_{s}, \\
\label{eq:dVb}
	-iV^{(1,0)2}_{2bubble} & = & {(-i \lambda)^2 \vphi \over 4} 
	\kappa^{4 \vep}	\left[ 
	\int {d^n k_1 \over (2 \pi)^n} \{ iD_{\beta}^{11}(k_1) \}^2 
	\int {d^n k_2 \over (2 \pi)^n} iD_{\beta}^{11}(k_2) 
	\right. \nonumber \\
	&& \left. - \int {d^n k_1 \over (2 \pi)^n} \{iD_{\beta}^{12}(k_1) \}^2
	\int {d^n k_2 \over (2 \pi)^n} iD_{\beta}^{22}(k_2) \right]
	+ counterterms, \\
\label{eq:defV10c}
	-iV^{(1,0)2}_c & = & -i {\lambda \over 2} (-iB_1 \mu^2 -iC_1 \vphi^2)
	\kappa^{2 \vep} \int {d^n k \over (2 \pi)^n} 
	\{iD_{\beta}^{11}(k) \}^2 - \{iD_{\beta}^{12}(k) \}^2 \nonumber \\
	&& -i {C_1 \vphi^2 \over 2} \kappa^{2 \vep} 
	\int {d^n k \over (2 \pi)^n} iD_{\beta}^{11}(k), \\
\label{eq:dsetsun}
	-i V^{(1,0)2}_{s} & = & {(-i \lambda \vphi)^3 \over 4} \kappa^{6 \vep}
	\int {d^n k_1 \over (2 \pi)^n} {d^n k_2 \over (2 \pi)^n}
	[ \{ iD_{\beta}^{11}(k_1)\}^2 iD_{\beta}^{11}(k_2) 
	iD_{\beta}^{11}(k_1 + k_2) \nonumber \\
	&& - 2 iD_{\beta}^{11}(k_1) iD_{\beta}^{12}(k_1) iD_{\beta}^{12}(k_2) 
	iD_{\beta}^{12}(k_1 + k_2) \nonumber \\
	&& + \{ iD_{\beta}^{12}(k_1)\}^2 iD_{\beta}^{22}(k_2) 
	iD_{\beta}^{22}(k_1 + k_2) ] \nonumber \\
	&& + {(-i \lambda \vphi) (-i \lambda) \over 6} \kappa^{4 \vep}
	\int {d^n k_1 \over (2 \pi)^n} {d^n k_2 \over (2 \pi)^n}
	[ iD_{\beta}^{11}(k_1) iD_{\beta}^{11}(k_2) 
	iD_{\beta}^{11}(k_1 + k_2) \nonumber \\
	&& - iD_{\beta}^{12}(k_1) iD_{\beta}^{12}(k_2) 
	iD_{\beta}^{12}(k_1 + k_2)] 
	+ counterterms, 
\eeq
 where,
\beq
\label{eq:Tpropagator}
	iD_{\beta}^{ab} (k) & = &
	\left( \begin{array}{cc}
	{i \over k^2 - \mu^2 + i \vep} & 0 \\[0.3cm]
	0 & {-i \over k^2 - \mu^2 - i \vep}
	\end{array} \right) \\
	&& + 2 \pi \delta(k^2 - \mu^2)
	\left( \begin{array}{cc}
	{1 \over e^{\beta |k_0|}-1} 
	& {- e^{- \beta |k_0|/2} \over 1 - e^{- \beta |k_0|}}\\[0.3cm]
	{- e^{- \beta |k_0|/2} \over 1 - e^{- \beta |k_0|}}
	& {1 \over e^{\beta |k_0|}-1} 
	\end{array} \right), \nonumber
\eeq
\beq
\label{eq:somedef}
	M^2 = \mu^2 + {\lambda \over 2} \vphi^2, \ \ \ 
	n_{B}(E) = (e^{\beta E}-1)^{-1}, \ \ \ 
	\cosh^2 \theta = { 1 \over 1- e^{- \beta |k_0|}},
\eeq
 and $\kappa$ is the renormalization point. The ``$(1,0)1$'' in $V^{(1,0)1}$ 
 indicates a function which has a one-point type 1 external field 
 and a zero-point type 2 external field at the one-loop level.
 $-iV^{(1,0)2}_{2bubble}$,
 $-iV^{(1,0)2}_c$ and $-i V^{(1,0)2}_{s}$ are contributed from 
 Figs.\ref{fig:2tado} (a),(b),\ref{fig:2tado} (c)-(e) and
 Fig.\ref{fig:setsun}, respectively.

 The above quantities are then integrated over $\vphi$, yielding the following:
\beq
\label{eq:V1}
	V^1 & = & -{1 \over (4 \pi)^2}{M^4 \over 4}
	({3 \over 2}-\ln{M^2 \over \kappa^2}) + \int^{\infty}_0 
	{dk \over (2 \pi)^2} {2 \over \beta}k^2 \ln(1-e^{-\beta E}), \\
\label{eq:2bubble}
	V^2_{2bubble} & = & 
	{\lambda \over 2} K_t^2 - {\lambda M^2 \over 4} C_f, \\
\label{eq:marubatsu}
	V^2_c & = & {\lambda \over 4}(\mu^2 + 
	{3 \over 2} \lambda \vphi^2) C_f, \\
\label{eq:setsun}
	V^2_{s} & = & {\lambda^2 \vphi^2 \over 4} \left[ {M^2 \over (4 \pi)^4} 
	\{ a + (\gamma - \ln{4 \pi} - 3 + \ln{M^2 \over \kappa^2}) 
	\ln{M^2 \over \kappa^2} \} \right. \nonumber \\
	&& \left. + {1 \over 2 (2 \pi)^2} (\ln{M^2 \over \kappa^2} 
	-2 + {\pi \over \sqrt{3}})I_t + I_l \right] \nonumber \\
	& \equiv & {\lambda^2 \vphi^2 \over 4} S_s, 
\eeq
 where,
\beq
	E & = & \sqrt{k^2 + M^2}, \\
\label{eq:defKt}
	K_t & = & {M^2 \over 2 (4 \pi)^2} (1- \ln{M^2 \over \kappa^2}) - I_t\\
\label{eq:defCf}
	C_f & = & {M^2 \over (4 \pi)^4} \{ d - (\gamma - \ln 4 \pi -1 
	+ {1 \over 2} \ln {M^2 \over \kappa^2}) \ln {M^2 \over \kappa^2} \},\\
\label{eq:defd}
	d  & = & -{1 \over 2} (\gamma^2 - 2 \gamma + 2 + {\pi^2 \over 6})
		+ (\gamma -1 - {1 \over 2} \ln 4 \pi) \ln 4 \pi
		\simeq -5.68496, \\
\label{eq:defa}
	a & = & {\pi^2 \over 12}+{9 \over 2}+{1 \over 2}(\gamma - \ln 4 \pi)
	(\gamma - \ln 4 \pi -3) \nonumber \\
	&& + \int^1_0 dx \int^1_o dy - \ln y (\ln \alpha 
	- {(1-y) \beta \over \alpha}) \nonumber \\
	& \simeq & 10.16186 - 2.17195 = 7.98891, \\
	\alpha & = & -(y-1 + {y \over x^2 -x}), \ \ 
	\beta  = -(1 + {1 \over x^2 -x}), \\
\label{eq:defIt}
	I_t & = & \int^{\infty}_0 {dk \over (2 \pi)^2}{k^2 n_{B}(E) \over E},\\
\label{eq:defIl}
	I_l & = & \int^{\infty}_0 {dk_1 dk_2 \over (2 \pi)^4}
	{k_1 k_2 n_{B}(E_1) n_{B}(E_2) \over 2 E_1 E_2}
	\ln \left|{4 E_1^2 E_2^2 - (M^2 + 2 k_1 k_2)^2 \over
	4 E_1^2 E_2^2 - (M^2 - 2 k_1 k_2)^2 }\right|,
\eeq
 and $\gamma$ is the Euler constant. 

\begin{figure}[ht]
\centerline{
    \epsfysize=2cm 
    \epsfbox{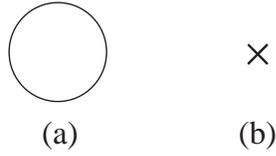}
}
     \caption{One-loop effective potential diagrams.}
\label{fig:V1}
\end{figure}
\begin{figure}[ht]
\centerline{
    \epsfysize=2.2cm 
    \epsfbox{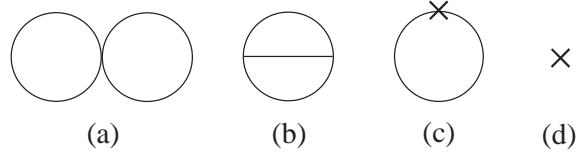}
}
     \caption{Two-loop effective potential diagrams.}
\label{fig:V2}
\end{figure}

\section{OPT Equations}
\label{app:fullV}

 Here, we summarize the calculations of the effective potential and 
 its derivatives in the full OPT.
\begin{figure}[ht]
\centerline{
    \epsfysize=3.5cm 
    \epsfbox{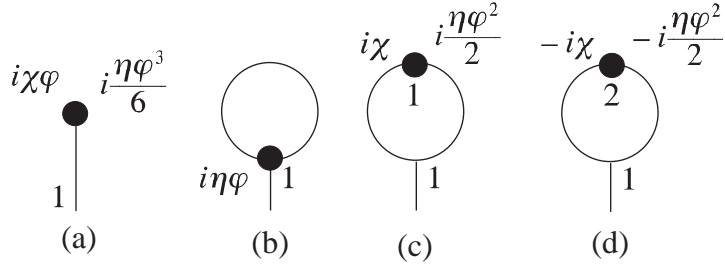}
}
     \caption{(a) Additional $O(\delta)$ contributions to the 1-point 
	function in the full OPT.
	(b)-(d) Additional $O(\delta^2)$ contributions 
	to the 1-point function.}
\label{fig:atado}
\end{figure}
\begin{figure}[ht]
\centerline{
    \epsfysize=2.8cm 
    \epsfbox{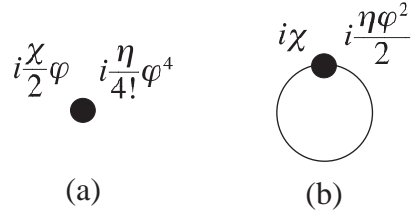}
}
     \caption{(a) Additional $O(\delta)$ contributions 
	to the effective potential in the full OPT.
	(b) Additional $O(\delta^2)$ contributions.}
\label{fig:aloop}
\end{figure}

 The two-loop effective potential in the full optimized perturbation 
 theory reads
\beq
\label{eq:fullV}
	V & = & V^{0} + V^{\delta} + V^{\delta^2},\\
	V^{0}  & = & {1 \over 2} m^2 \vphi^2 +{g \over 4!} \vphi^4,\\
	V^{\delta}  & = & -{1 \over 2} \chi \vphi^2 -{\eta \over 4!} \vphi^4
		-{1 \over (4 \pi)^2}{M^4 \over 4}
		({3 \over 2}-\ln{M^2 \over \kappa^2}) \nonumber \\
		&& + \int^{\infty}_0 {dk \over (2 \pi)^2}
		{2 \over \beta}k^2 \ln(1-e^{-\beta E}), \\
	V^{\delta^2} & = & (\chi + {\eta \vphi^2 \over 2} 
		+ {g \over 2} K_t) K_t + {g^2 \vphi^2 \over 4} T_f,
\eeq
 where,
\beq
\begin{array}{lcccl}
	\mu^2   & = & m^2 - (m^2 -\mu^2) & = & m^2 - \chi,\\
	\lambda & = & g - (g - \lambda)  & = & g - \eta,
\end{array}
\eeq
\beq
\label{eq:defKtf}
	K_t & = & {M^2 \over 2 (4 \pi)^2} (1- \ln{M^2 \over \kappa^2}) - I_t,\\
\label{eq:defTf}
	T_f & \equiv & S_s + C_f \nonumber \\
	& = & {M^2 \over (4 \pi)^4} \{ c + (- 2 + {1 \over 2}
	\ln{M^2 \over \kappa^2}) \ln{M^2 \over \kappa^2} \} \nonumber \\
	&& + {1 \over 2 (2 \pi)^2} (\ln{M^2 \over \kappa^2} -2 
	+ {\pi \over \sqrt{3}})I_t + I_l,
\eeq
\beq
\label{eq:defItf}
	I_t & = & \int^{\infty}_0 {dk \over (2 \pi)^2}{k^2 n_{B}(E) \over E},\\
\label{eq:defIlf}
	I_l & = & \int^{\infty}_0 {dk_1 dk_2 \over (2 \pi)^4}
		{k_1 k_2 n_{B}(E_1) n_{B}(E_2) \over 2 E_1 E_2}
		\ln \left|{4 E_1^2 E_2^2 - (M^2 + 2 k_1 k_2)^2 \over
		4 E_1^2 E_2^2 - (M^2 - 2 k_1 k_2)^2 }\right|,
\eeq
\beq
	M^2 = m^2 + {g \over 2} \vphi^2,\ \ 
	E   = \sqrt{k^2 + M^2},\ \ 
	n_{B}(E) = (e^{\beta E}-1)^{-1},
\eeq
\beq
\label{eq:defc}
	c & = & {1 \over 2} (7 -\gamma + \ln 4 \pi) - \int^1_0 dx \int^1_0 
		dy \ln y (\ln \alpha - {(1-y) \beta \over \alpha}) \nonumber \\
	& \simeq & 2.30495, \\
	\alpha & = & -(y-1 + {y \over x^2 -x}),\ \ 
	\beta  = -(1 + {1 \over x^2 -x}),
\eeq
 and $\gamma = 0.5772 \cdots$.

\subsection{For PMS conditions}
\label{app:PMS}

 Here, we give equations needed to study the PMS conditions.  
 In the following, 
 we set $g = \lambda$, so $\eta = 0$ in Eq.(\ref{eq:fullV}):
\beq
\label{eq:dV1f}
	{\d V \over \d \vphi} & = & {\d V^0 \over \d \vphi}
	+ {\d V^{\delta} \over \d \vphi} 
	+ {\d V^{\delta^2} \over \d \vphi},\\
	{\d V^0 \over \d \vphi}
	& = & \vphi \left[ m^2 + {\lambda \over 6} \vphi^2 \right],\\
	{\d V^{\delta} \over \d \vphi} 
	& = &  \vphi \left[ - \chi - \lambda K_t \right], \\
	{\d V^{\delta^2} \over \d \vphi}
	& = & \vphi \left[ \lambda (\chi + \lambda K_t) {\d K_t \over \d M^2}
	+ {\lambda^2 \over 2} 
	(T_f + {\lambda \vphi^2 \over 2} {\d T_f \over \d M^2}) \right].
\eeq
 The partial derivative with respect to $m^2$ is found to be
\beq
\label{eq:dvdm2}
	{\d V \over \d m^2} & = & {\d V^0 \over \d m^2}
	+ {\d V^{\delta} \over \d m^2} 
	+ {\d V^{\delta^2} \over \d m^2},\\
	{\d V^0 \over \d m^2}
	& = & {1 \over 2} \vphi^2,\\
	{\d V^{\delta} \over \d m^2} 
	& = & - K_t - {1 \over 2} \vphi^2, \\
	{\d V^{\delta^2} \over \d m^2} 
	& = & K_t + ( \chi + \lambda K_t) {\d K_t \over \d M^2} 
	+ {\lambda^2 \vphi^2 \over 4}{\d T_f \over \d M^2}.
\eeq
 Also, when $\vphi = 0$, (\ref{eq:dvdm2}) read
\beq
\label{eq:fPMSml}
	\left. {\d V \over \d m^2} \right|_{\vphi =0} = 
	( \chi + \lambda K_t) {\d K_t \over \d M^2}. 
\eeq

\subsection{For FAC conditions}
\label{app:FAC}

 The second derivative of the effective potential with respect 
 to $\vphi$ for the FAC conditions is found to be
\beq
\label{eq:dV2f}
	{\d^2 V \over \d \vphi^2} & = & {\d^2 V^0 \over \d \vphi^2}
	+ {\d^2 V^{\delta} \over \d \vphi^2} 
	+ {\d^2 V^{\delta^2} \over \d \vphi^2},\\
	{\d^2 V^0 \over \d \vphi^2} 
	& = & m^2 + {\lambda \over 2} \vphi^2,\\
	{\d^2 V^{\delta} \over \d \vphi^2}
	& = & -\chi - \lambda K_t
	- \lambda^2 \vphi^2 {\d K_t \over \d M^2},\\
	{\d^2 V^{\delta^2} \over \d \vphi^2}
	& = & \lambda {\d K_t \over \d M^2} (\chi + \lambda K_t
	+ \lambda^2 \vphi^2 {\d K_t \over \d M^2})
	+ \lambda^2 \vphi^2 {\d^2 K_t \over \d (M^2)^2}
	(\chi + \lambda K_t) \nonumber \\
	&& + {\lambda^2 \over 2} (T_f + {5 \over 2} \lambda \vphi^2
	{\d T_f \over \d M^2} + {1 \over 2} \lambda^2 \vphi^4
	{\d^2 T_f \over \d (M^2)^2}).
\eeq

\subsection{High $T$ forms}

 Here, we give the high $T$ ($\beta M \rightarrow 0$) forms 
 of Eq.(\ref{eq:defKtf}), derivatives of Eq.(\ref{eq:defKtf}) and
 Eq.(\ref{eq:defIlf}). We find
\beq
	K_t & \rightarrow & -{T^2 \over 24} + {MT \over 8 \pi}
	+ {M^2 \over 2 (4 \pi)^2}(1 - \ln{T^2 \over \kappa^2})
	- {M^2 \over (4 \pi)^2}( \ln 4 \pi - \gamma + {1 \over 2}),
	\hspace{0.4cm} \\
	{\d K_t \over \d M^2} & \rightarrow & 
	{T \over 16 \pi M} + {1 \over 2 (4 \pi)^2}(1 - \ln{T^2 \over \kappa^2})
	- {1 \over (4 \pi)^2}( \ln 4 \pi - \gamma + {1 \over 2}),\\
	{\d^2 K_t \over \d (M^2)^2} & \rightarrow & 
	- {T \over 32 \pi M^3},\\
	I_l & \rightarrow & 
	{T^2 \over 12 (4 \pi)^2}(\ln{M^2 \over T^2} + 3.48871).
\eeq


\end{document}